\documentclass[a4paper,fleqn,usenatbib]{mnras}

\usepackage{savesym}
\usepackage{amsmath}
\savesymbol{iint}
\usepackage{txfonts}
\restoresymbol{TXF}{iint}

\usepackage[T1]{fontenc}
\usepackage{ae,aecompl}

\usepackage{graphicx}	
\usepackage{amsmath}	
\usepackage{amssymb}	
\usepackage{natbib}
\usepackage{caption,subcaption,tikz}
\tikzset{boximg/.style={remember picture,red,thick,draw,inner sep=0pt,outer sep=0pt}}

\usepackage{float}
\definecolor{edits}{RGB}{255,127,0}
\usepackage{soul}
\sethlcolor{yellow}
\soulregister\citet7
\soulregister\citep7
\soulregister\citealp7
\soulregister\ref7
\soulregister\url7
\usepackage{breakcites}

\title[Runaway Be stars]{On the kinematics of a runaway Be star population}

\author[Boubert \& Evans]{
D. Boubert,$^{1}$\thanks{E-mail: d.boubert,nwe@ast.cam.ac.uk}
and N. W. Evans$^{1}$
\\
$^{1}$Institute of Astronomy, University of Cambridge, Madingley Road, Cambridge CB3 0HA, UK\\
}

\date{Accepted XXX. Received YYY; in original form ZZZ}

\pubyear{2017}

\begin{document}
\label{firstpage}
\pagerange{\pageref{firstpage}--\pageref{lastpage}}
\maketitle

\begin{abstract}
    We explore the hypothesis that B type emission-line stars (Be
    stars) have their origin in mass-transfer binaries by measuring
    the fraction of runaway Be stars. We assemble the largest-to-date
    catalogue of 632 Be stars with 6D kinematics, exploiting the
    precise astrometry of the Tycho-Gaia Astrometric Solution (TGAS)
    from the first Gaia Data Release. Using binary stellar evolution
    simulations, we make predictions for the runaway and equatorial
    rotation velocities of a runaway Be star population. Accounting
    for observational biases, we calculate that if all classical Be
    stars originated through mass transfer in binaries, then $17.5\%$
    of the Be stars in our catalogue should be runaways. The remaining
    82.5\% should be in binaries with subdwarfs, white dwarfs or
    neutron stars, because those systems either remained bound
    post-supernova or avoided the supernova entirely. Using a Bayesian
    methodology, we compare the hypothesis that each Be star in our
    catalogue is a runaway to the null hypothesis that it is a member
    of the Milky Way disc. We find that $13.1^{+2.6}_{-2.4}\%$ of the
    Be stars in our catalogue are runaways, and identify a subset of
    40 high-probability runaways. We argue that deficiencies in our
    understanding of binary stellar evolution, as well as the
    degeneracy between velocity dispersion and number of runaway
    stars, can explain the slightly lower runaway fraction. We thus
    conclude that all Be stars could be explained by an origin in
    mass-transfer binaries. This conclusion is testable with the
    second Gaia data release (DR2).
\end{abstract}




\section{Introduction}
\label{sec:introduction}
Around 17\% of B type stars in the Milky Way show emission lines in
their spectra \citep{zorec_critical_1997} and are referred to as the
classical Be stars. This designation is historically restricted to
  stars of luminosity class V-III, as proposed by
  \citet{jaschek_be_1981}. They are believed to originate when a
rapidly rotating B star forms a decretion disc
\citep{rivinius_classical_2013}, albeit for unknown reasons.  The
rapid rotation of the central star minimises the velocity that
material in the atmosphere must reach in order to form a Keplerian
decretion disc. Both pulsations and magnetic fields have been
suggested as ways to launch material into a disc \citep[see e.g.,][and
  references therein]{rivinius_classical_2013}. An alternative is that
the Be phenomenon is related to binary
interactions. \citet{harmanec_emission-line_1987} suggested that the
disc is formed of material lost by a Roche-lobe filling companion.
However, there is little evidence for such a mass-losing companion for
most Be stars. \citet{pols_formation_1991} proposed an alternative
mechanism, now known as the post-mass-transfer model, whereby mass is
transferred to a star by a Roche-lobe filling companion and the
angular momentum carried by this material spins the star up to close
to critical velocity. This model requires that a star spinning at or
near critical velocity can spontaneously form a decretion disc through
an unknown mechanism, which could be the pulsations or magnetic fields
proposed for single Be star channels.

Evidence for the mass transfer hypothesis comes from the large number
of Be star in binaries. These should be Be+NS (neutron star) binaries
if the primary explodes as a supernova. If the primary is sufficiently
stripped, it may avoid a supernova explosion and be present as a white
dwarf or subdwarf star. There are five known Be+sdO binaries
\citep{gies_hubble_1998,peters_detection_2008,peters_far-ultraviolet_2013,peters_hot_2016,wang_detection_2017}
and 28 confirmed Be+NS X-ray binaries \citep{reig_be/x-ray_2011}.

Recently, \citet{boubert_binary_2017} conducted a search of ten nearby
Galactic supernova remnants for the runaway former companion of the
progenitor star. Four candidates in four remnants were identified, one
of which, BD+50 3188 in remnant HB 21, was found to be a Be
star. \citet{boubert_binary_2017} argued that this was circumstantial
evidence both for the runaway candidacy of the star and for the
post-mass-transfer model in general.

There are several other recent works which lend support to the
post-mass-transfer model. \citet{chernyakova_study_2017} studied the
Be-X-ray binary LSI +61$^{\circ}$ 303, which has an eccentricity
$e>0.5$. They argue that the superorbital variability seen in this
system is due to the compact remnant passing through different regions
of the Be star disc and so probing material of varying densities. One
origin for this large eccentricity could be that the compact remnant
was formed in a supernova and received a large natal kick. Another way
to produce an eccentric binary is through capture in a dense stellar
system, but LSI +61$^{\circ}$ 303 does not appear to be situated in
such an environment. \citet{gonzalez-galan_multi-wavelength_2018}
studied SXP 1062, a Be-X-ray binary located in the Small Magellanic
Cloud, which is likely associated with the supernova remnant MCSNR
J0127-7332. SXP 1062 is the first Be-X-ray binary that is likely
residing in its parental supernova remnant. \citet{naze__2017}
examined $\pi$ Aquarii, a 14 solar mass Be star orbited by a 3 solar
mass main-sequence companion at 1 AU separation. This system could
arise if it was originally a triple and the inner binary merged to
form what is now the Be star. The merging could have occurred during
common envelope evolution, the compact remnant of the binary might
have been kicked into the companion or the inner binary could have
remained bound post-supernova and merged at a later point.

Runaway stars are thought to form through one of two channels. In
  the Binary Supernova Scenario (BSS, \citealp{blaauw_origin_1961}), a
  star is ejected from a binary system by the supernova of its more
  massive companion. In the Dynamical Ejection Scenario (DES,
  \citealp{poveda_run-away_1967}), 3- or 4-body encounters during the
  collapse of a young star cluster cause the ejection of one or more
  of the stars. It is not known definitively which of these two
  mechanisms dominates, but the BSS is thought to be more likely to
  due to the ubiquity of binary systems among massive stars
  (e.g.\citealp{branch_supernova_2017}). If Be stars predominantly
originate through the post-mass-transfer route, then a significant
fraction should also be runaway stars. \citet{rinehart_single_2000}
constructed a sample of 5756 B and 129 Be stars in the Hipparcos
dataset \citep{esa_hipparcos_1997} and used them to look for
differences between the peculiar velocity distribution of B and Be
stars, concluding that the distributions are identical to within
$1\sigma$ and thus that the post-mass-transfer model is not supported
by the data. \citet{berger_search_2001} searched for high-velocity Be
stars by cross-matching an existing catalogue of Be stars with the
Hipparcos catalogue \citep{esa_hipparcos_1997}. They only considered
stars with published radial velocities brighter than $V \approx 9$ to
ensure inclusion in Hipparcos. \citet{berger_search_2001} classified
all stars with a peculiar space velocity greater than
$40\;\mathrm{km}\;\mathrm{s}^{-1}$ as high-velocity and hypothesised
that they originate either with a supernova that disrupted the
progenitor binary or with binary-binary dynamical interactions in
young clusters. 23 of the 344 Be stars in their sample have peculiar
space velocities greater than
$40\;\mathrm{km}\;\mathrm{s}^{-1}$. Because there is substantial mass
transfer before the supernova, we expect that most systems remain bound
after the supernova and so these numbers could be consistent with the
post-mass-transfer model.

The impetus for looking at these problems anew comes from the Gaia
satellite, which was launched on 19th December
2013~\citep{gaia_collaboration_gaia_2016-1}. This is a successor
satellite to Hipparcos and is monitoring all objects brighter than $V
\approx 20$ over a period of 5 years. It is providing magnitudes,
parallaxes, proper motions and broad band colours for over a billion
stars. It is a valuable new resource for problems at the intersection
between stellar evolution and stellar dynamics. In this paper, we
use the Tycho-Gaia Astrometric Solution (TGAS), which is one of the
catalogues comprising the Gaia Data Release
1~\citep{gaia_collaboration_gaia_2016}. It uses data from
Tycho-2~\citep{hog_tycho-2_2000} to provide a 30 year baseline for
astrometric calculations and provides parallaxes and proper motions on
over 2 million stars mostly within $\approx 1$ kpc of the Sun.

Here, we assemble the largest catalogue of Be stars with full
  six-dimensional kinematics in Sec. 2, exploiting the remarkable
  precision of the TGAS astrometric catalogue.  We predict the runaway
  velocity distribution for Be stars based on simulations of binary
  star evolution in Sec. 3, accounting for the observational bias of
  the catalogue. In Sec. 4, we formulate a Bayesian approach to our
  problem of estimation of the fraction of Be stars that are runaways
  from binary supernova. We present our main result that around $13\%$
  of the Be stars in our catalogue are consistent with being runaway
  stars in Sec. 5, and list the 40 most probable examples.  Finally,
  in Sec. 6, we discuss whether our results are consistent with all Be
  stars originating through the post-mass-transfer channel, and show
  that we can explain the puzzling lack of Be stars in the
  high-latitude sample of runaway B stars of
  \citet{martin_origins_2006}. Finally, in the Conclusion, we argue
  that we can resolve the true number of Be star runaways by applying
  the Bayesian methodology introduced in Sec. \ref{sec:bayesian} to
  the second data release of the Gaia satellite.

\section{Data}
\label{sec:data}

Extracting constraints on the origin of Be stars requires
well-measured radial velocities, proper motions and distances. We
consider three datasets which satisfy this requirement. The first is
from \citet{berger_search_2001}, the second is a cross-match between
the Be Star Spectra database (BeSS) with SIMBAD and the Tycho-Gaia
Astrometric Solution (TGAS), and the third is a cross-match of the Be
catalogue of \citet{hou_catalog_2016} from the Large Sky Area
Multi-Object Fibre Spectroscopic Telescope (LAMOST) with TGAS. We then
combine and clean these datasets by removing any duplicates, as well
as stars with anomalously high radial velocities. Whilst the fraction
of runaways among the Be stars is unknown, the fraction of runaway B
stars has long been established to be $\approx 2.5$\% for stars of
type B0--0.5 and declining to $\approx 1.5$\% for stars of type B1--B5
\citep{blaauw_origin_1961}. It is important to note that there is a difference in terminology between theory and observations when it comes to runaway stars. Runaways fractions are calculated observationally by classifying any star with a peculiar velocity greater than $40\;\mathrm{km}\;\mathrm{s}^{-1}$ to be a runaway star, while theoretically any star ejected from a binary by the supernova explosion of the companion can be called a runaway star with the majority having ejection velocities below this value.


\subsection{The Berger \& Gies sample}
\label{sec:bergerdata}

\begin{figure}
	\includegraphics[scale=0.55,trim = 4mm 4mm 0mm 3mm, clip]{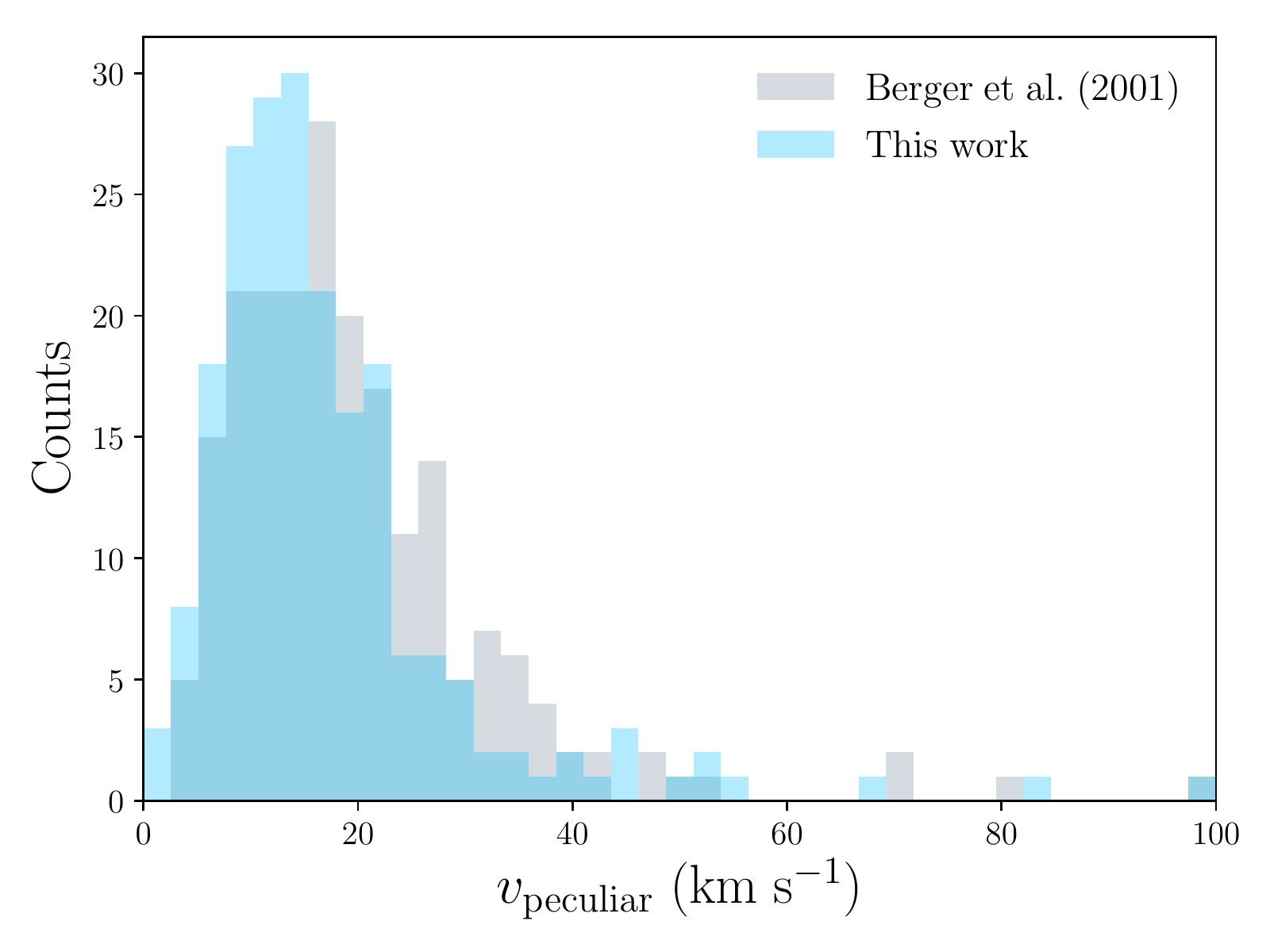}
	\caption{Median peculiar velocities relative to the Milky Way
          disc of an identical sample of stars using the kinematics
          from either \citet{berger_search_2001} or the catalogue
          assembled in this work. Thanks to TGAS, the kinematics in
          our catalogue are more accurate and this produces a
          significant tightening in the distribution.}
	\label{fig:bergervpec}
\end{figure}

The \citet{berger_search_2001} sample contains 344 stars. We first
obtain positions and updated radial velocities by querying SIMBAD
using the identifiers listed by \citet{berger_search_2001}. Note that
the identifier of the thirteenth star in this sample, CSI+6101449,
does not have a cross-match in SIMBAD, but querying Vizier returned
the valid identifier HD 10664. For 262 stars, the error on the radial
velocity recorded in SIMBAD is smaller than in
\citet{berger_search_2001}. For 320 stars, the radial velocities are
consistent within the reported error in the two catalogues.  For the
other 24 stars, we checked the source of the radial velocity in
SIMBAD. Only for HD 120991, where the radial velocity reference was
\citet{wilson_general_1953}, did we judge the SIMBAD radial velocity
to be less reliable than the \citet{berger_search_2001} value. In all
other cases, we took whichever radial velocity measurement had the
smallest error. The majority of the new radial velocities come from
\citet{gontcharov_pulkovo_2006} and
\citet{kharchenko_astrophysical_2007}. Using the positions, we
cross-match with TGAS and obtain more accurate parallaxes and proper
motions for 163 of the stars.  In Fig.~\ref{fig:bergervpec}, we
illustrate the effect of the updated measurements on the peculiar
velocity distribution. Note that the peculiar velocities for
\citet{berger_search_2001} are taken from that work and use a solar
radius $R_{\odot}=8.5\;\mathrm{kpc}$, a local circular speed
$v_{\mathrm{disc}}=220\;\mathrm{km}\;\mathrm{s}^{-1}$ and a solar
peculiar velocity
$(U,V,W)_{\odot}=(10.0,5.25,7.17)\;\mathrm{km}\;\mathrm{s}^{-1}$. For
the peculiar velocities in this work, we use the more recent values
stated in Section \ref{sec:bayesian}. 

\subsection{Be Star Spectra database}
\label{sec:bessdata}

The Be Star Spectra database (BeSS) claims to be a complete catalogue
of classical Be stars and Herbig Ae/Be stars. It contains information
on 2265 Be stars and reports radial velocities for 856 and radial
velocity errors for 759. After collating this information, we queried
SIMBAD to obtain radial velocities, parallaxes and proper
motions. SIMBAD provides the proper motion errors in terms of an error
ellipse with a major axis $A$, minor axis $B$ and position angle
$P$. We convert these to uncertainties in the individual components
following the recommended
method\footnote{\url{http://simbad.u-strasbg.fr/Pages/guide/errell.htx}}
and neglect the implied covariances. We then carry out a
$1\;\mathrm{arcsec}$ nearest neighbour cross-match with TGAS to obtain
parallaxes and proper motions. To select whether to use the radial
velocity from BeSS or SIMBAD, we pick whichever had the smallest
associated error. If the two radial velocities differ by more than
$1\sigma$ with respect to their errors added in quadrature, we look at
the origin of both measurements and in all cases prefer the radial
velocity quoted through SIMBAD. In almost all cases, it was possible
to trace the velocity quoted in BeSS back to the General Catalogue of
Stellar Radial Velocities and its revision
\citep{wilson_general_1953,evans_revision_1967}. These velocities have
been superseded.

\subsection{The LAMOST Sample}
\label{sec:houdata}

LAMOST is carrying out a 5 year spectroscopic survey of 10 million
Milky Way stars in the Northern hemisphere down to
$20.5\;\mathrm{mag}$. \citet{hou_catalog_2016} presented a catalogue
of 10,436 early-type emission-line stars which are in LAMOST.  This
catalogue is available as a value-added catalogue of LAMOST
DR2\footnote{\url{http://dr2.lamost.org/doc/vac}}. We carry out a
$1\;\mathrm{arcsec}$ nearest neighbour cross-match with the LAMOST DR2
stellar catalogue to obtain the reported radial velocities with errors
and with TGAS to obtain the parallax $\omega$ and proper-motions
$(\mu_{\alpha\ast},\mu_{\delta})$. This results in a total of 12
stars. We add $6.76\;\mathrm{km}\;\mathrm{s}^{-1}$ to correct for the
offset found between LAMOST and SDSS SEGUE radial velocities
\citep{jing_kinematics_2016}.

 \begin{figure}
	\includegraphics[scale=0.55,trim = 4mm 4mm 0mm 3mm, clip]{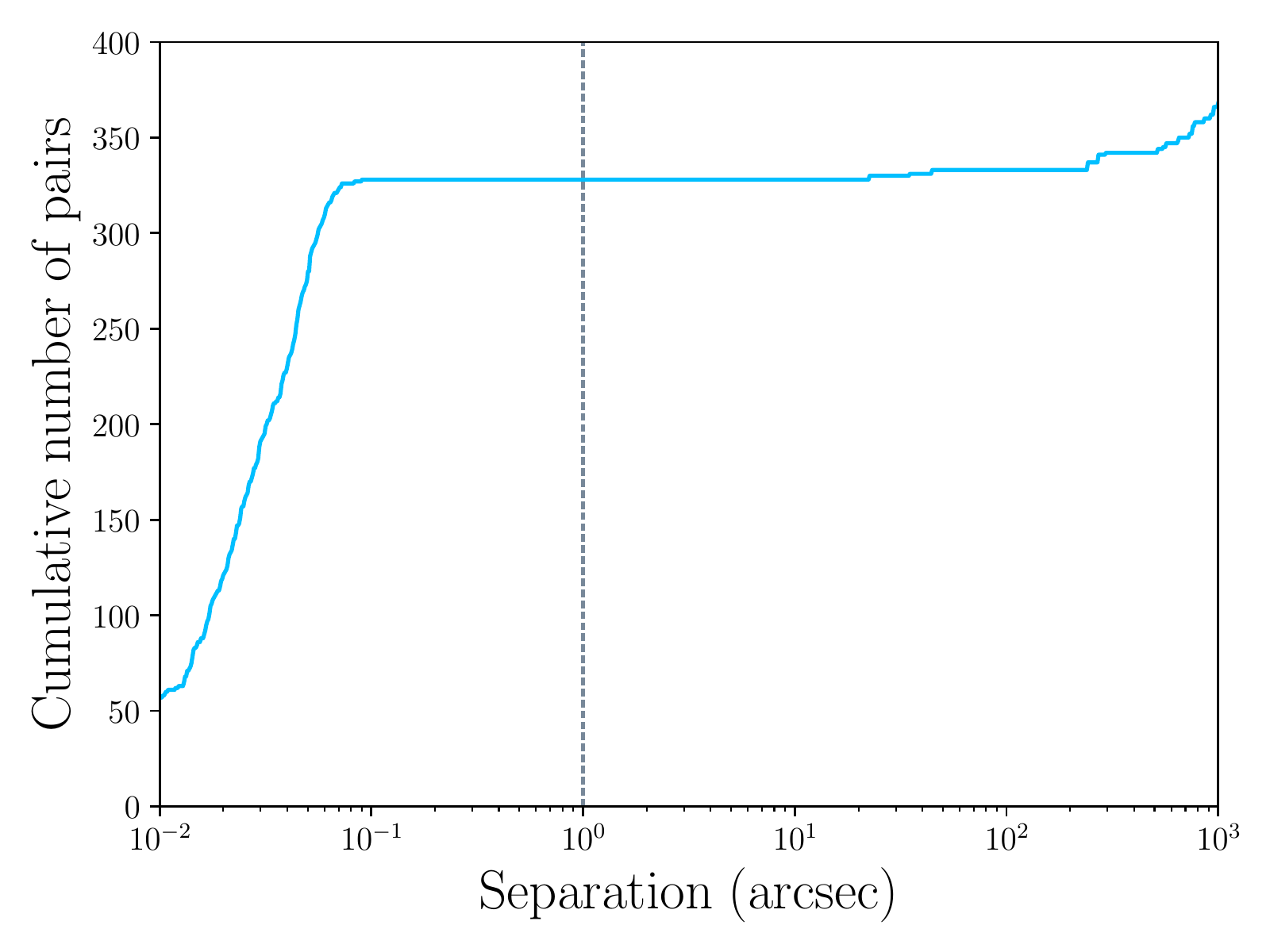}
	\caption{Cumulative number of pairs of stars with a separation
          less than a given bound. The dashed line is our criterion
          for a duplicate star and it cleanly divides duplicate stars
          from true close stellar pairs.}
	\label{fig:duplicates}
\end{figure}

\subsection{The Combined Catalogue}
\label{sec:threesources}

We combine all three sources into one dataset. There is a large
cross-over between the \citet{berger_search_2001} and BeSS
datasets. We project each star onto the unit sphere and use those 3D
positions to construct a $k$-dimensional tree. We call a duplicate any
pair of stars in the combined dataset that are separated by less than
$1\;\mathrm{arcsec}$. There are 328 pairs that meet this criterion. As
Fig. \ref{fig:duplicates} shows, there are no additional pairs in the
range $1\mathrm{-}10\;\mathrm{arcsec}$ and thus there is no blurring
between the duplicate population and the population of close stars
that exist in binaries or dense clusters.

\begin{figure}
	\begin{tikzpicture}
	\node[anchor=south west,inner sep=0] at (0,0) {\includegraphics[scale=0.39,trim = 0mm 12mm 3mm 16mm,
		clip]{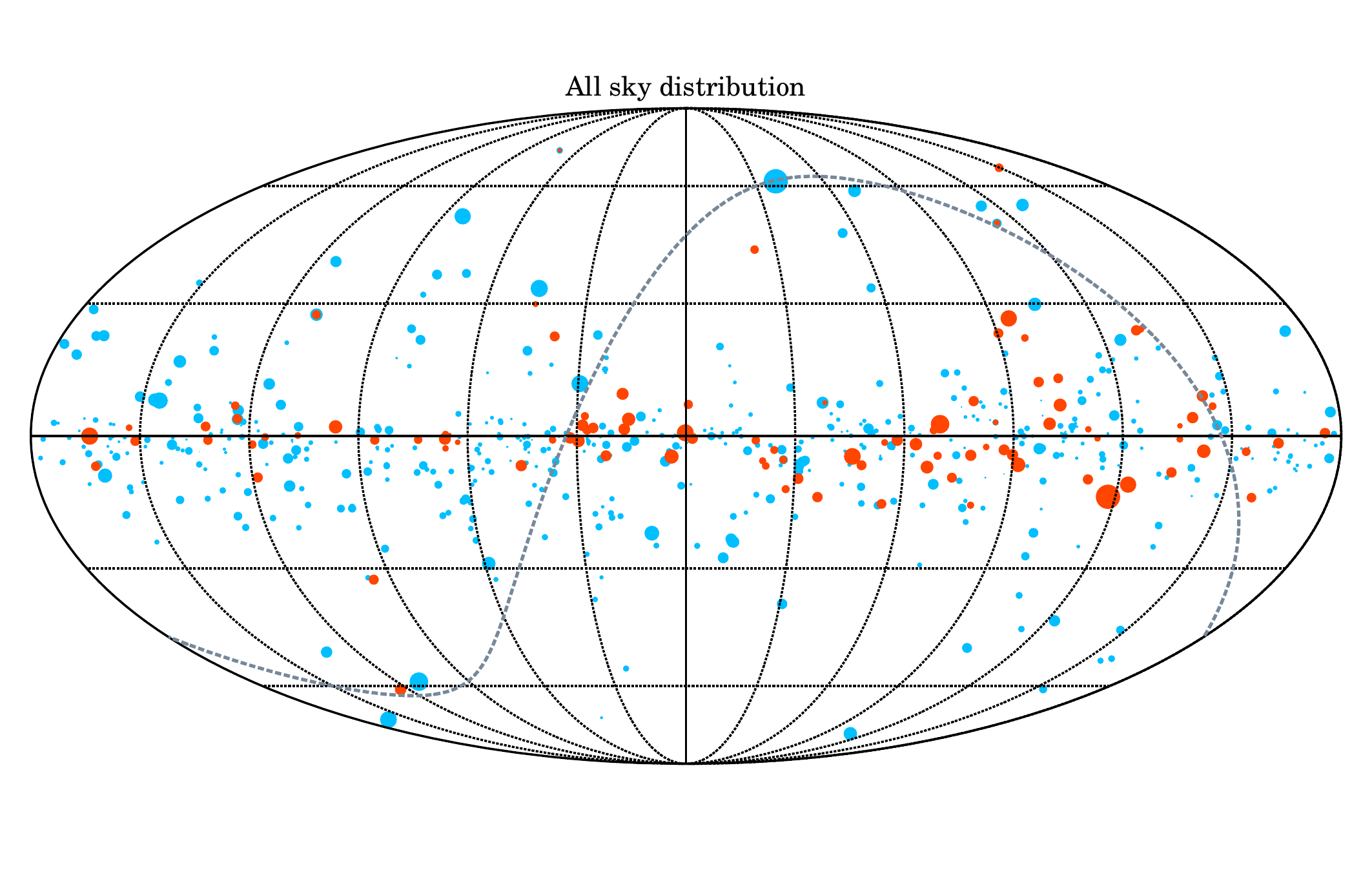}};
	\node[] at (4.25,-0.05) {\footnotesize Galactic longitude};
	\node[rotate=90] at (-0.02,2.15) {\footnotesize Galactic latitude};
	\end{tikzpicture}
	\caption{Distribution of Be stars in our combined
            catalogue across the sky. The size of each point is
            proportional to the parallax. The Galactic centre is at
            the centre of this image and the celestial equator is
            shown as a grey dashed line. A subset of the points are
            coloured orange to indicate stars which do not have
            an associated spectral type and luminosity class. Such
            stars are preferentially found in the Southern
            hemisphere.}
	\label{fig:lb}
\end{figure}

\begin{figure}
	\includegraphics[scale=0.55,trim = 4mm 4mm 0mm 3mm,
          clip]{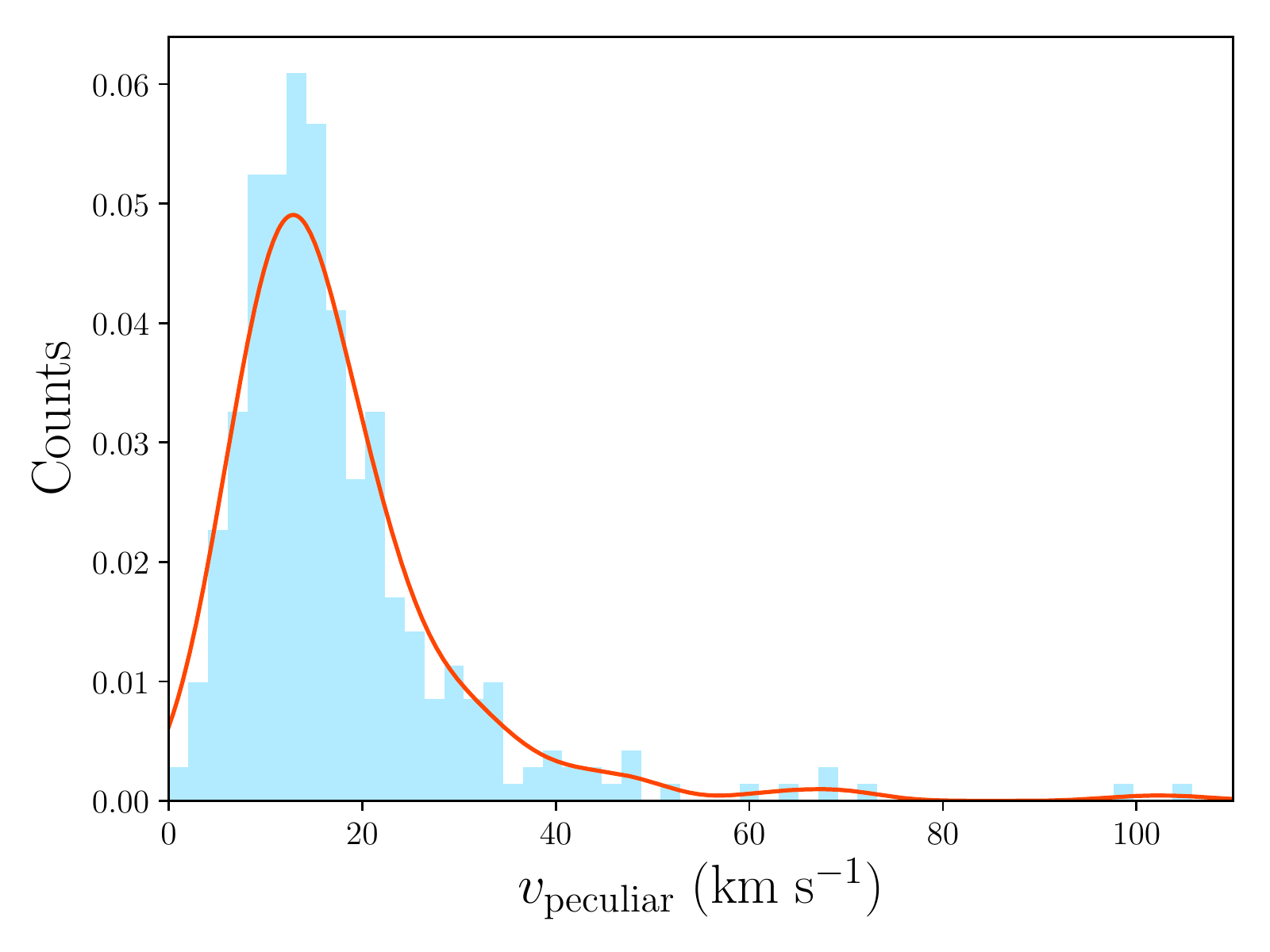}
	\caption{Median peculiar velocity distribution of Be stars in
          our combined catalogue. Overplotted is a kernel density
          estimate.}
	\label{fig:vpec}
\end{figure}

We find that three of the stars in our sample have unusually high
radial velocities; CPD-32 2038 at
$912.11\pm97.922\;\mathrm{km}\;\mathrm{s}^{-1}$, HD 152979 at
$-588.45\pm51.38\;\mathrm{km}\;\mathrm{s}^{-1}$, and HD 165783 at
$-507.353\pm66.777\;\mathrm{km}\;\mathrm{s}^{-1}$. These velocities
are from RAVE DR3 and DR4. In all three cases, the RAVE velocity
measurements are the only ones that exist for these stars. Since all
other stars in our sample have velocities
$|v_{\mathrm{rad}}|<200\;\mathrm{km}\;\mathrm{s}^{-1}$, these are very
likely to be spurious measurements. We remove all three stars from our
sample. There are a further two stars, HD 306989 and CD-61 4751, whose
only reported radial velocity from \citet{reed_catalog_2003} has no
associated radial velocity error and so these are also removed. This
leaves us with a final sample of 632 stars. Their all-sky distribution
in Galactic coordinates and median peculiar velocity distribution are
shown in Figs.~\ref{fig:lb} and ~\ref{fig:vpec} respectively.

The Be star sample described above is a biased subset of all Be
  stars, partly due to the diversity of sources from which it has been
  drawn. The most significant bias is over-representation of
  early-type, giant stars, because our sample is flux-limited rather
  than volume-limited. We quantify this bias by using the spectral
  types and luminosity classes of the sample obtained from
  SIMBAD. Among the 632 stars, 599 have an associated spectral type
  and 532 also have a luminosity class. Multiple stars in the
  catalogue have a luminosity class of I or II, which contradicts the
  standard definition of a Be star and thus suggests that there is a degree of uncertainty in the luminosity classes. Similarly, there must be uncertainty in the stellar types because 38 stars are classified as Be stars but do not have a B spectral type. We choose to trust the Be star identification over the spectral type or luminosity class and thus include all 632 stars in our analyses in later sections.
  
\begin{figure}
	\includegraphics[scale=0.485,trim = 3mm 7mm 34mm 6mm,
	clip]{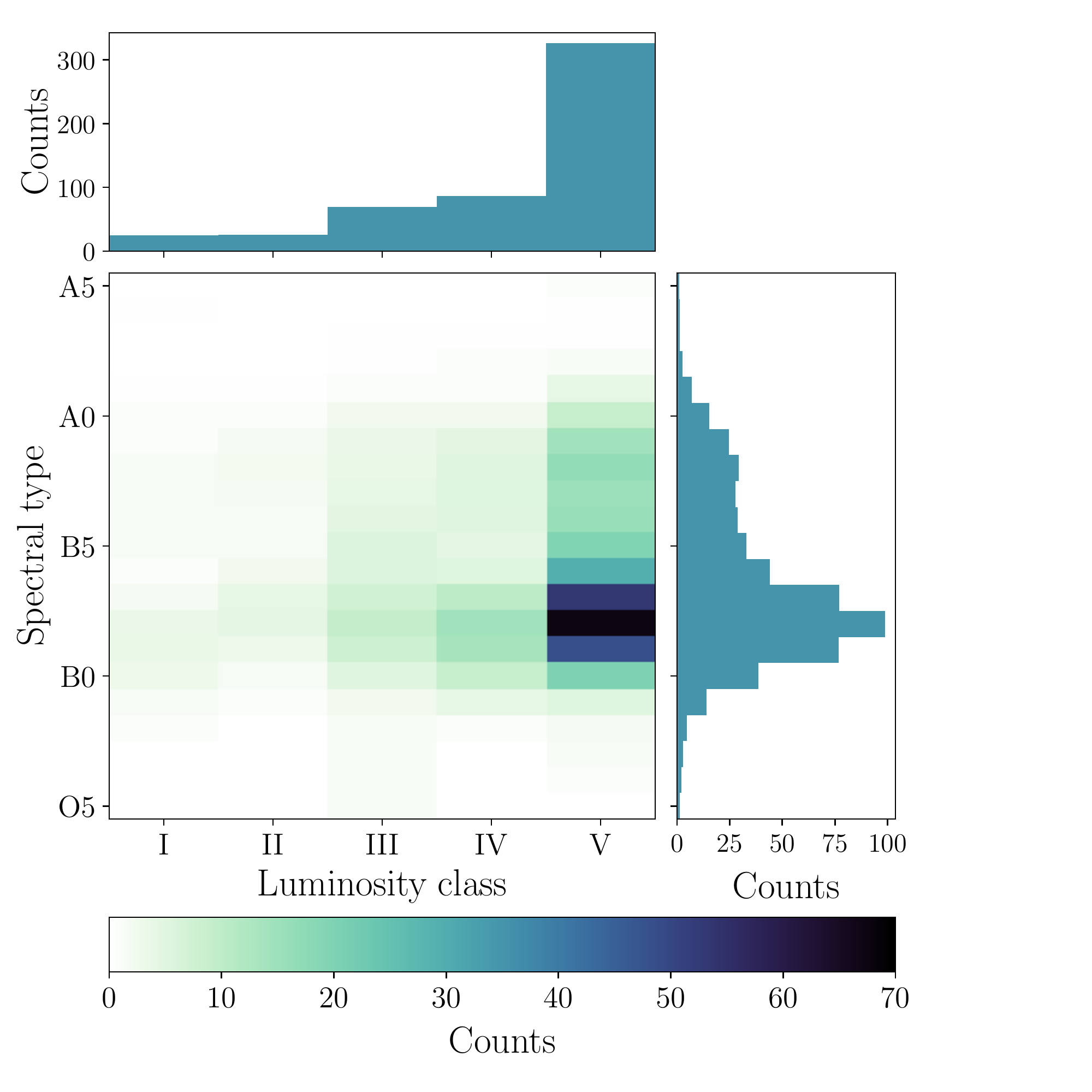}
	\caption{Distribution of the 532 Be stars which have both
            spectral types and luminosity classes. The counts have
            been convolved with a Gaussian kernel of width 0.2322 to
            smooth out shot noise. There is a clear observational bias
            towards earlier types, while 38 stars are classified as Be
            stars but do not have a B spectral type.}
	\label{fig:speclum}
\end{figure}

  We bin the stars in the two
  dimensional space of their spectral type
  $\{\mathrm{O}0,\mathrm{O}1,...,\mathrm{G}0\}$ and luminosity class
  $\{\mathrm{I},\mathrm{II},\mathrm{III},\mathrm{IV},\mathrm{V}\}$. Note
  that the spectral type or luminosity class can indicate a range of
  possible classifications. In this case, we divide up the
  contribution of that star to the histogram between all possible
  combinations of spectral type and luminosity class (i.e. a star
  whose classification was B1/2Iab/II would contribute a quarter to
  each of the bins B1I, B1II, B2I and B2II). We then apply a Kernel
  Density Estimation to this histogram with a bandwidth of 0.2322
  (estimated using the rule of \citealp{scott_multivariate_2015} and
  assuming that each bin has unit length and width) which acts to
  smooth the histogram and reduce shot noise. This smoothed
  distribution is shown in Fig. \ref{fig:speclum}. We make the
  assumption that this distribution adequately describes the selection
  function of the entire 632 stars. This assumption is well-motivated;
  the distribution of spectral types does not change significantly if
  we include the 67 stars which have spectral types, but not luminosity
  classes. Notably, Fig. \ref{fig:lb} demonstrates that the stars
  which do not have these classifications are found preferentially in
  the Southern hemisphere, presumably due to an historical bias in the
  geographical distribution of telescopes. This bias further motivates
  the inclusion of all 632 stars in order to avoid spatial bias in
  our catalogue.

\section{The Post-Mass-Transfer Channel for Be Stars}
\label{sec:binaryc}

The aim of this section is to use binary stellar population
  synthesis to predict the population properties of Be stars.

\subsection{Be stars from binary population synthesis}

Be stars produced through the post-mass-transfer route have interacted
with their companion in the past. The runaway velocity is strongly
dependent on their past evolution and is imprinted in their present
day velocities. We model the evolution of binaries across a grid in
parameter space using the {\sc binary\_c} population-nucleosynthesis
framework~\footnote{We use version 2.0pre28, SVN 5018.}
\citep{izzard_new_2004,izzard_population_2006,izzard_population_2009}. This
code is based on the binary-star evolution ({\sc bse}) algorithm of
\citet{hurley_evolution_2002} expanded to incorporate nucleosynthesis,
wind-Roche-lobe-overflow \citep{abate_wind_2013,abate_modelling_2015},
stellar rotation \citep{de_mink_rotation_2013}, accurate stellar
lifetimes of massive stars \citep{schneider_ages_2014}, dynamical
effects from asymmetric supernovae \citep{tauris_runaway_1998}, an
improved algorithm describing the rate of Roche-lobe overflow
\citep{claeys_theoretical_2014}, and core-collapse supernovae
\citep{zapartas_delay-time_2017}. In particular, we take our black
hole remnant masses from \citet{spera_mass_2015} and use a fit to the
simulations of \citet{liu_interaction_2015} to determine the impulse
imparted by the supernova ejecta on the companion.  Grids of stars are
modelled using the {\sc binary\_grid2} module to explore the
binary-star parameter space in primary mass $M_{1}$, secondary mass
$M_{2}$ (or equivalently mass ratio $q$) and orbital period
$P_{\mathrm{orb}}$. The grid variables have the ranges
\begin{align}
1.0 \leq M_1 / M_{\sun} &\leq 80.0,\nonumber\\
0.1 \; M_{\sun}/M_1  \leq q &\leq 1,\\
-1.0 \leq \log_{10} (P_{\mathrm{orb}}/\mathrm{days}) &\leq 10.0.\nonumber
\end{align}
We assume the primary mass has the \mbox{\cite{kroupa_variation_2001}} IMF,
\begin{equation}
N(M_1)\propto
\begin{cases}
M_1^{-0.3}, & \mathrm{if}\ 0.01<M_1/M_{\sun}<0.08, \\
M_1^{-1.3}, & \mathrm{if}\ 0.08<M_1/M_{\sun}<0.5, \\
M_1^{-2.3}, & \mathrm{if}\ 0.5<M_1/M_{\sun}<80.0, \\
0, & \mathrm{otherwise.}
\end{cases}
\end{equation}
We assume a flat mass-ratio distribution for each system over the
range $0.1\;M_{\sun}/M_1<q<1$. We use the hybrid period distribution
from \citet{izzard_binary_2018} which gives the period
distribution as a function of primary mass and bridges the log-normal
distribution for low-mass stars \citep{duquennoy_multiplicity_1991}
and a power law \citep{sana_binary_2012} distribution for OB-type
stars. The grid is set at solar metallicity since Be stars have
short lifetimes of less than $1\;\mathrm{Gyr}$, despite rejuvenation
through mass-transfer extending the life of Be stars through the
post-mass-transfer route by up to several $100\;\mathrm{Myr}$. {\sc
  binary\_c} has previously been used to consider the origins of Be
stars by \citet{de_mink_rotation_2013} in their investigation of the
rotation rates of massive stars. They concluded that the 24.1\% of
their simulations that resulted in mass gain by the companion or a
merger with the primary is consistent with the $20-30\%$ of early B
type stars which are found to be Be stars \citep{zorec_critical_1997}.

Be stars are dwarf or giant B type stars with emission-lines in
  the spectra. However, the atmospheres of stars are not modelled in
detail by {\sc binary\_c}. We thus require a definition of a Be star
based on the dynamic properties of the star. The natural
  expression is in terms of the ratio $R_{\mathrm{eq}}$ of the
  equatorial velocity $v_{\mathrm{eq}}$ to the critical equatorial
  velocity for break-up
  $v_{\mathrm{eq,crit}}$. \citet{townsend_be-star_2004} note that,
  while the canonical value for this ratio is
  $R_{\mathrm{eq}}\simeq0.7-0.8$ based on measurements of the rotation
  of Be stars, this earlier work neglected the effect of equatorial
  gravity darkening. They postulate that this effect could lead to the
  Be star rotation rate being underestimated by tens of percent and
  that a value of $R_{\mathrm{eq}}\simeq0.95$ is not ruled out. This
  near-critical rotation is in better agreement with the scenario
  first described by \citet{struve_origin_1931}, in which material
  leaks out from the equator of a star spinning near breakup. To
  illustrate this point, \citet{townsend_be-star_2004} note that to
  launch material ballistically into orbit from the surface of a star
  spinning at $R_{\mathrm{eq}}\simeq0.7$ requires an additional
  $100\;\mathrm{km}\;\mathrm{s}^{-1}$. \citet{rivinius_classical_2013}
  reviewed the observational evidence and concluded that the minimum
  ratio for a B star to become a Be star is around 0.7 and that the
  lower limit of the mean ratio of the Be stars is around 0.8. For
  instance, \citet{rivinius_bright_2006} found
  $R_{\mathrm{eq}}=0.75\pm0.14$ as a lower limit without including
  gravity darkening.  For this work, we consider the range
  $R_{\mathrm{eq}}\in(0,1)$ but focus on the set of plausible values
  $R_{\mathrm{eq}}\in\{0.65,0.75,0.85,0.95\}$. Based on the
  considerations above and results presented in Section
  \ref{sec:obsbias}, we choose the fiducial value
  $R_{\mathrm{eq}}=0.85$ to use in Section \ref{sec:bayesian}
  onwards. The properties of the Be stars appear to be robust to the
  precise choice of $R_{\mathrm{eq}}$.

Another consideration is that, if a binary remains bound
  post-supernova, its centre of mass experiences a kick
  \citep{nelemans_constraints_1999}. This kick is not included at
  present in \textsc{binary\_c}. However, \citet{berger_search_2001}
  find that for a typical Be star scenario, the systematic kick is
  $8(\Delta M/\mathrm{M}_{\odot})\;\mathrm{km}\;\mathrm{s}^{-1}$
  where $\Delta M$ is the mass lost by the primary in the
  supernova. They further note that this mass ratio is of order unity
  and thus that high velocity runaway binaries containing Be stars are
  not expected in general. In agreement with this theoretical
  expectation, observations of Be star - X-ray binaries in the Galaxy
  find that they have low peculiar velocities of
  $15\pm6\;\mathrm{km}\;\mathrm{s}^{-1}$
  \citep{van_den_heuvel_origin_2000}, broadly similar to the velocity
  dispersions of a population of B stars containing no runaways
  (e.g. \citealp{aumer_kinematics_2009}).

\subsection{Properties of the simulated Be star population accounting for selection effects}
\label{sec:obsbias}

\begin{figure}
	\includegraphics[scale=0.60,trim = 6mm 4mm 0mm 2mm, clip]{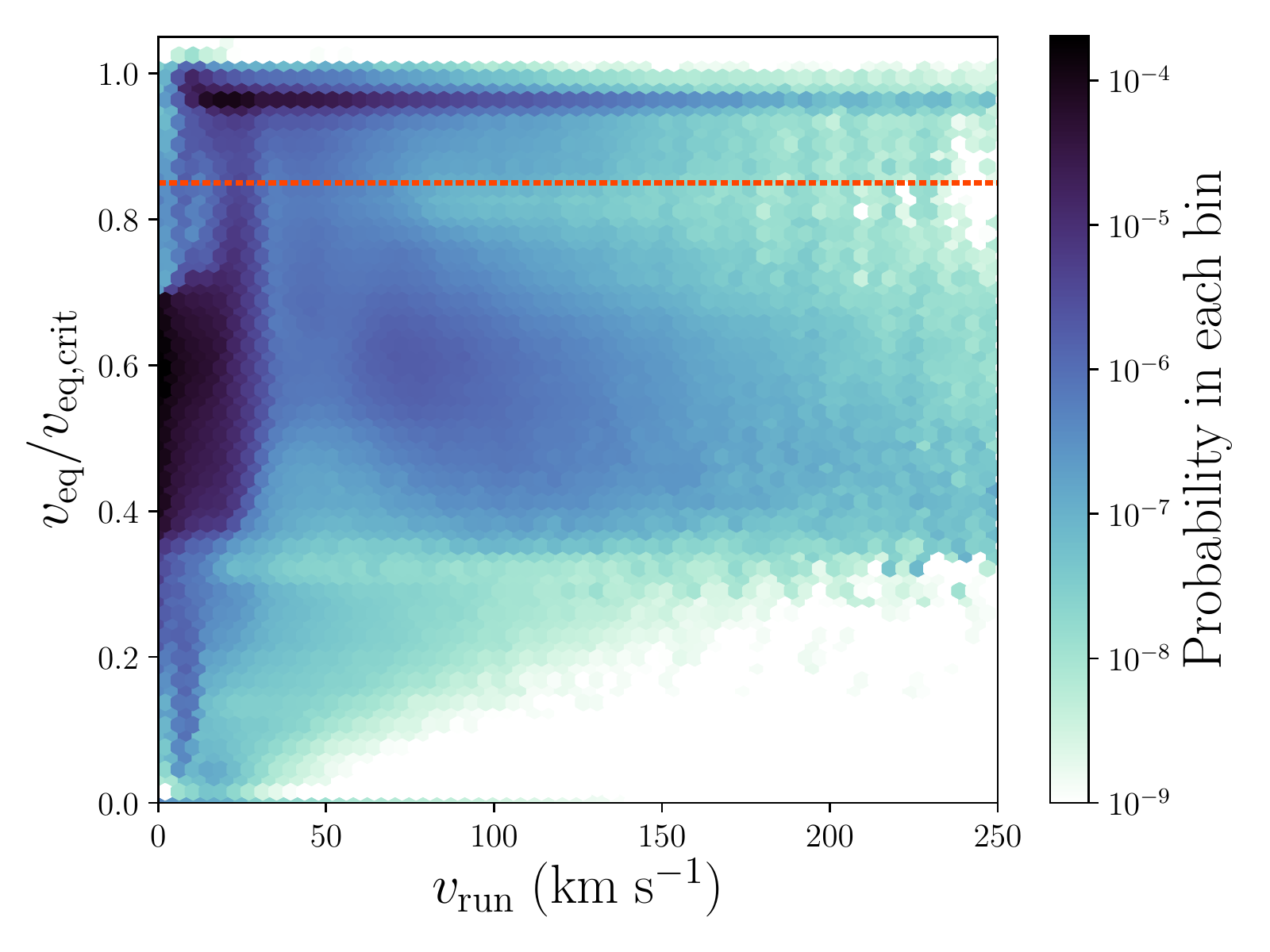}
	\caption{Predicted probability distribution for the runaway
          velocity and critical equatorial velocity ratio of runaway B
          stars. The dashed line indicates our fiducial value
          $R_{\mathrm{eq}}=0.85$.}
	\label{fig:vejveq}
\end{figure}

The relationship between equatorial rotation and runaway velocity
  is central to understanding the post-mass-transfer model for the Be
  star phenomenon. First, mass transfer from the primary to the
secondary shrinks the binary and accelerates the orbital velocity, and the orbital velocity is the principal contributor to the post-supernova runaway velocity. This mass transfer spins up the secondary, implying that high
velocity runaways will also be rapidly rotating. Second, two massive stars in a close orbit tidally
lock. This has the effect of breaking the relation between runaway
velocity and rotation velocity for close binaries, which correspond to
very fast runaways. This balance between mass transfer spinning up and
tidal locking spinning down companions is demonstrated by a plot of
runaway velocity $v_{\mathrm{run}}$ versus equatorial rotation
velocity $v_{\mathrm{eq}}$ for all the simulated, runaway B stars (see
Fig. \ref{fig:vejveq}), where there is a clear uptick in the
equatorial velocity of the population over the range $30\lesssim
v_{\mathrm{ej}} \lesssim 70 \; \mathrm{km}\;\mathrm{s}^{-1}$. The
structure of Fig.~\ref{fig:vejveq} is sensitive to the details of the
binary evolution simulation, but in broad terms stars start on the
left hand edge with a range of velocities. Some experience mass
transfer from the primary which both spins them up and increases the
orbit velocity. If the star is spun up to
$v_{\mathrm{eq}}/v_{\mathrm{eq,crit}}>1$, then material is lost from
the equator until it returns below the critical value. The star
remains as a Be star unless mass transfer from the primary shrinks the
orbit to the point where the timescale for tidal locking becomes
short, in which case the star is spun down. Note that the separation,
and hence runaway velocity, at which tidal locking becomes effective
is a function of the masses of both components. This is the reason for
the broad range of orbital velocities at which the companion is spun
down.

The predicted properties of Be stars discussed in the remainder of
this section assume a steady-state in which there has been a constant
star formation rate for longer than the age of the longest lived B
star. The implication is that these distributions are predictions for
the observable population rather than the stars produced in a single
starburst. At each timestep of the binary evolution, we check
  whether either component of the binary is a B star. If so, we
  tabulate both the properties of the binary and the probabilistic weight
  of the system given by the length of the timestep $\mathrm{d}t$
  multiplied by the probability of the progenitor binary
  $\operatorname{P}(M_1,q,P_{\mathrm{orb}})$.

The frequency of Be stars among the simulated B stars
  monotonically depends on our choice of $R_{\mathrm{eq}}$. Across
  spectral types B0-B9 and luminosity classes V-III, we find that the
  frequency of Be stars is 7.9\% for $R_{\mathrm{eq}}=0.65$ and
  declines to 1.9\% for $R_{\mathrm{eq}}=0.95$. These frequencies are
  lower than are observed
  (e.g. \citealp{zorec_critical_1997}). However, these predicted
  frequencies are sensitive to uncertain prescriptions for the birth
  rotation velocities and rotational evolution of the simulated
  stars. For instance, the birth rotation velocities in the fiducial
  model are given by an empirically-derived formula from
  \citet{hurley_comprehensive_2000}. If this prescription
  under-predicts the birth rotation velocities, then this will lead to
  a lower frequency of Be stars. We investigate this by setting the
  birth rotation velocity of each star to be the critical rotation
  velocity. In this extreme case, we find a Be star frequency of
  81.7\% for $R_{\mathrm{eq}}=0.65$ which declines to 3.3\% for
  $R_{\mathrm{eq}}=0.95$. Based on considerations in the previous
  section, a value $R_{\mathrm{eq}}=0.85$ seems most plausible, in
  which case our fiducial model gives a frequency of 3.1\% while the
  extreme model gives 25.5\%. These numbers bracket the observed
  frequency of Be stars and thus we conclude that an adjustment to the
  prescriptions for the birth rotation velocities and rotational
  evolution could bring our prediction in line with
  observations. While the extreme model boosts the abundance of Be
  stars by making an unphysical assumption, it does not substantially
  alter the properties of the resulting Be star population. Thus the
  predictions made for the properties of Be stars are robust to the
  adjustments necessary to replicate the observed Be star frequency.

Mass transfer can spin the companion up and cause the Be star
  phenomenon, but it also decreases the orbital separation and thus
  increases the velocity of the subsequent runaway Be star. One
  consequence is that, in rough terms, the longer the mass transfer
  continues, the more rapidly rotating the companion star and the
  faster the runaway velocity, implying that the frequency and
  velocity distribution of runaway stars amongst the Be star
  population depend on $R_{\mathrm{eq}}$. Additionally, the
  distribution of runaway star velocities is conditional on the mass
  of the star (for instance, see Fig. 2 in
  \citealp{boubert_binary_2017}) and this should hold for Be stars. An
  observable proxy for mass is the spectral type of the star and thus
  we can re-phrase the velocity distributions as being conditional on
  the spectral type. However, we established in Section \ref{sec:data}
  that our catalogue of Be stars is biased towards early-type Be
  stars. The implication is that the prediction for the runaway
  frequency and velocity distribution of the Be stars obtained through
  simulation is not directly comparable with our Be star catalogue. We
  account for this selection bias by re-weighting the simulated Be star
  population to match the observed luminosity class and spectral type
  distribution. Note that each choice of $R_{\mathrm{eq}}$ defines a
  different Be star population, and thus the re-weighting is carried
  out for each value of this parameter.
	
\begin{figure}
	\begin{subfigure}{0.5\textwidth}
		\includegraphics[scale=0.54,trim = 3mm 2mm 4mm 3mm, clip]{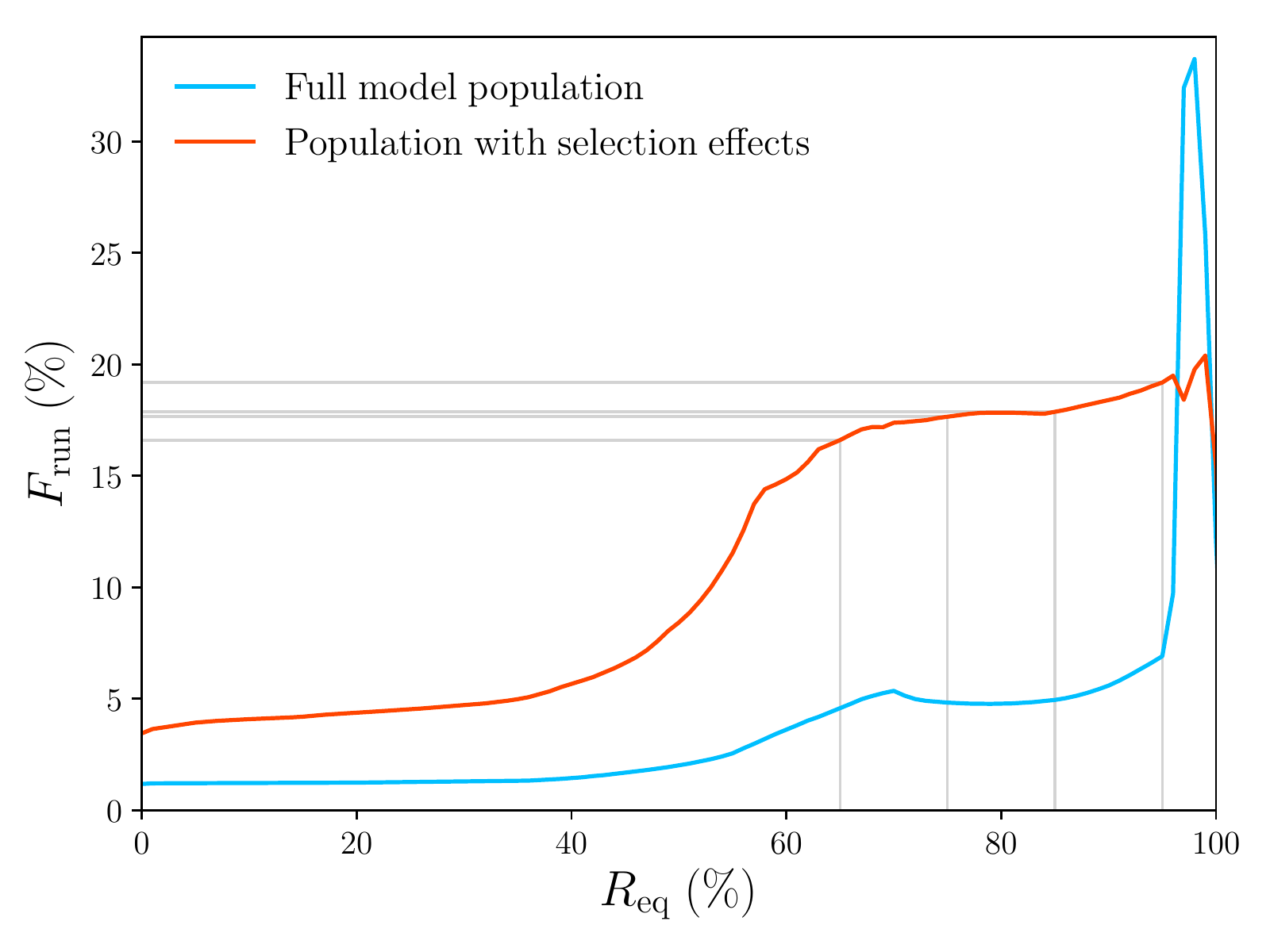}
		\caption{Runaway fraction as a function of $R_{\mathrm{eq}}$}
		\label{fig:runfracduetobias}
	\end{subfigure}
	\begin{subfigure}{0.5\textwidth}
		\includegraphics[scale=0.58,trim = 3mm 0mm 0mm 5mm, clip]{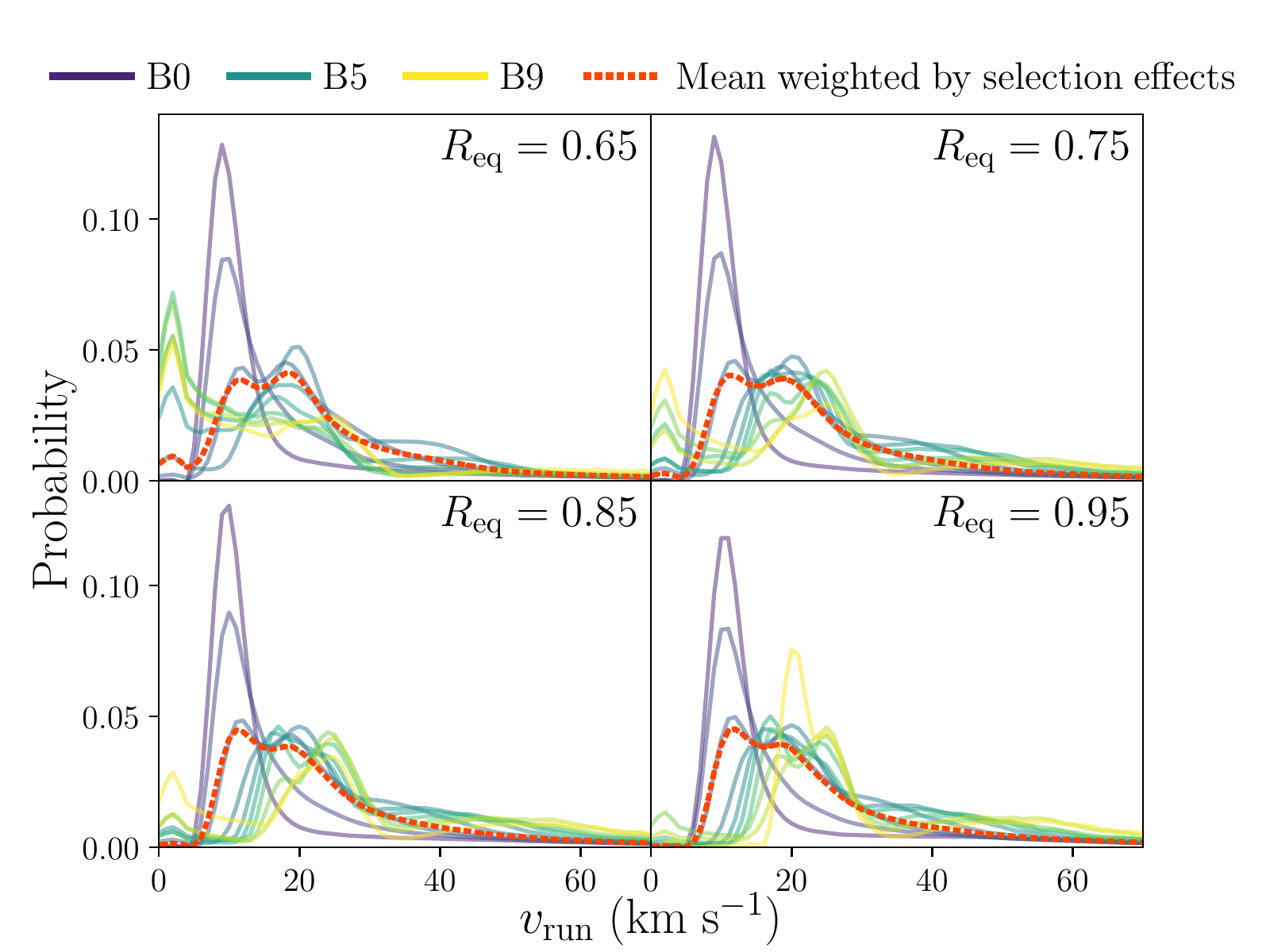}
		\caption{Velocity distribution of the runaway Be stars as a function of $R_{\mathrm{eq}}$}
		\label{fig:obsrunvel}
	\end{subfigure}
	\caption{Both selection effects and the
            uncertainty in the choice of $R_{\mathrm{eq}}$ (the
            minimum fraction of the critical rotation velocity for a
            star to be classed as a Be star) can lead to dramatically
            different simulated populations of runaway Be
            stars. \textbf{Top:} The runaway fraction among Be stars
            as a function of $R_{\mathrm{eq}}$ with and without
            accounting for the observational selection effect. The grey lines
            indicate the resulting runaway fraction associated with
            each choice of $R_{\mathrm{eq}}$ shown in the lower
            panel. \textbf{Bottom:} The choice of $R_{\mathrm{eq}}$
            changes the velocity distribution of the runaway Be stars
            as a function of spectral type. In each panel, we show the
            velocity distribution for each spectral type from B0 to
            B9, in addition to the mean accounting for the selection effect. Note
            that as $R_{\mathrm{eq}}$ increases the effect of Poisson
            noise also increases because we are defining fewer stars
            to be Be stars.}
	\label{fig:obsrunaway}
\end{figure}

\begin{figure}
	\begin{subfigure}{0.5\textwidth}
		\includegraphics[scale=0.58,trim = 0mm 0mm 5mm 12mm, clip]{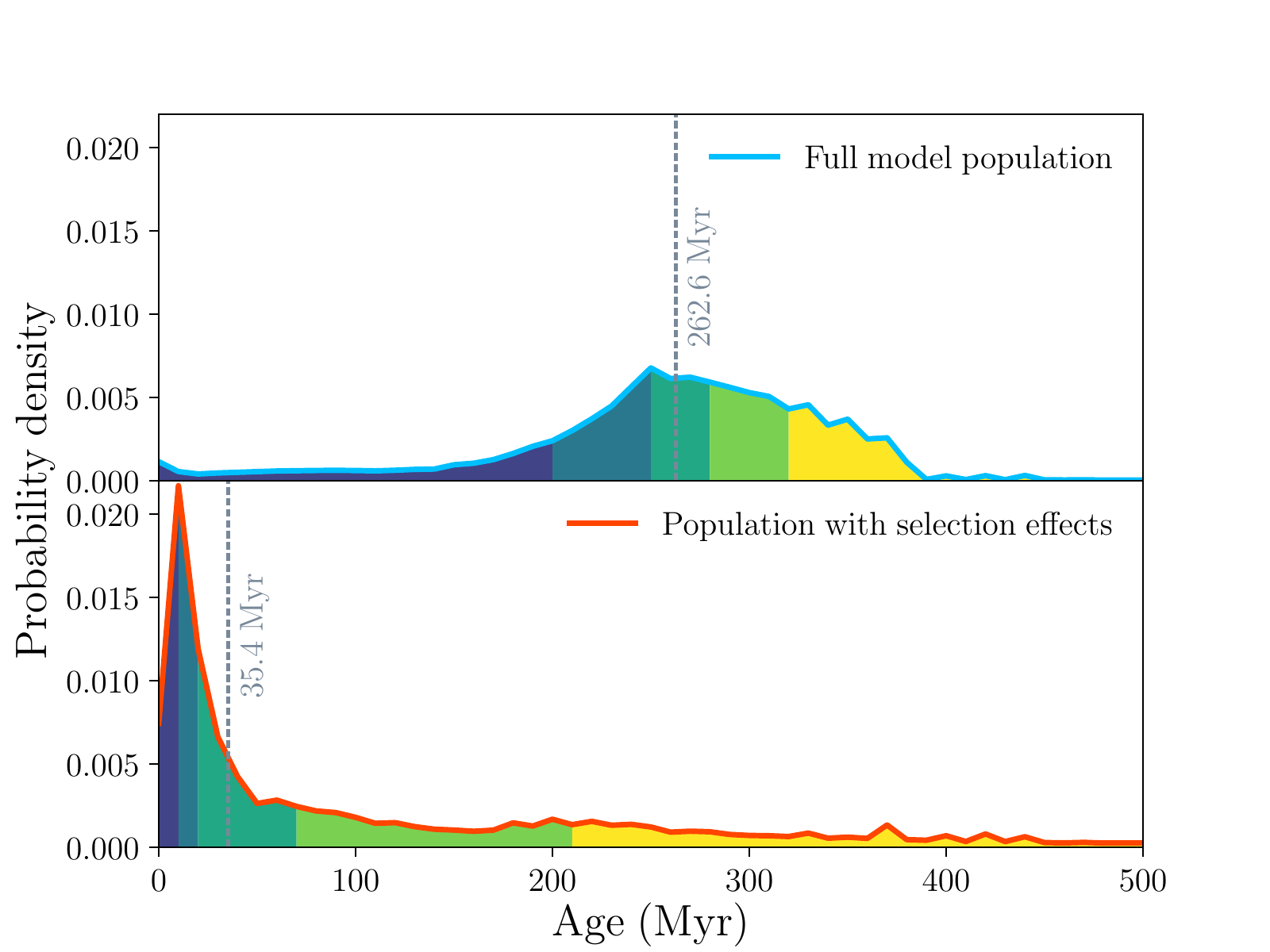}
		\caption{Age distribution of the simulated Be stars}
		\label{fig:obstime}
	\end{subfigure}
	\begin{subfigure}{0.5\textwidth}
		\includegraphics[scale=0.59,trim = 10mm 36mm 0mm 12mm, clip]{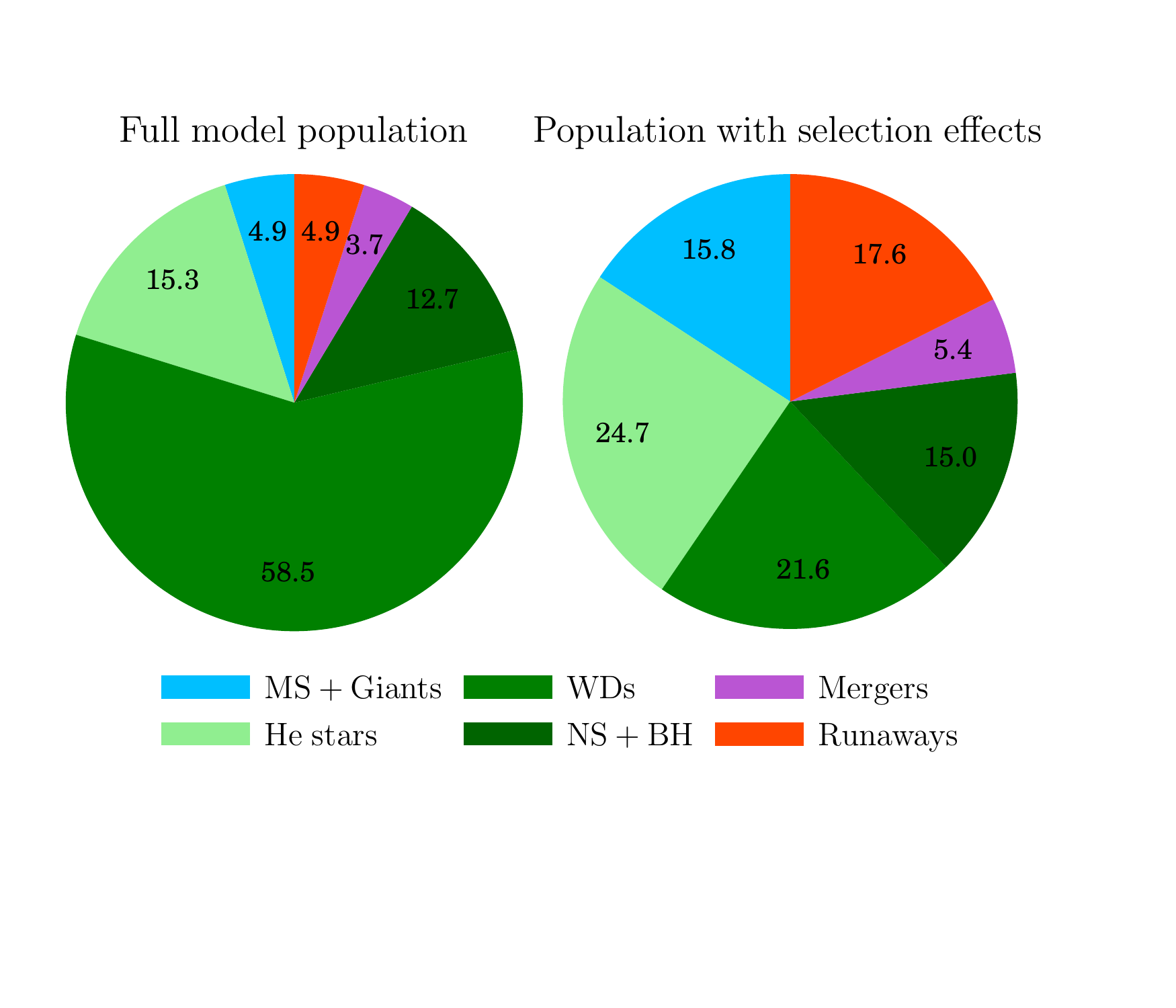}
		\caption{\% of Be stars where the binary is intact,
                  merged or split}
		\label{fig:obspartners}
	\end{subfigure}
	\caption{The inclusion of observational biases
            dramatically changes the predicted population of Be stars,
            under the assumption of
            $R_{\mathrm{eq}}=0.85$. \textbf{Top:} The two panels show
            the distribution of ages of the Be stars with and without
            accounting for the observational bias towards early-type
            Be stars. The grey dashed lines indicate the median age
            and the coloured regions each account for 20\% of the
            probability. \textbf{Bottom:} Each pie chart gives the
            proportions of the Be stars for which the progenitor
            binary is intact, merged or split. If the binary is intact
            then the stellar type of the companion is given. MS = Main
            Sequence, WDs = White Dwarf, NS = Neutron Star, BH = Black
            Hole.}
	\label{fig:obsall}
\end{figure}

Weighting the simulation outcomes by the observed luminosity class
  and spectral type distribution has a large effect both on the runaway
  star sub-population (Fig. \ref{fig:obsrunaway}) and on the Be
  star population in general (Fig. \ref{fig:obsall}). We refer to our Be star population from the synthesis as our full model population, while that with all known selection effects included will be referred to as the model population with selection effects.
\begin{enumerate}
	\item The predicted frequency of runaway stars amongst the model
          Be star population is increased, as shown in
          Fig.~\ref{fig:runfracduetobias}. Earlier stellar types tend
          to have a greater runaway fraction
          (e.g. \citealp{blaauw_origin_1961}) and the re-weighting
          increases the weight given to early-type Be
          stars. Fig.~\ref{fig:runfracduetobias} shows that
          choices of larger $R_{\mathrm{eq}}$ are correlated with
          greater runaway fractions. One possible reason is that stars
          which are closer to critical-rotation have experienced more
          mass-transfer, and thus tend to have a more massive partner
          whose core-collapse supernova would more easily disrupt the
          binary.
	\item The re-weighting alters the predicted runaway
          velocity distribution by increasing the contribution of
          early-type Be stars
          (Fig. \ref{fig:obsrunvel}). Interestingly, the
          mean accounting for selection effects is almost independent of the
          choice of $R_{\mathrm{eq}}$, making this mean robust to our
          fiducial choice of $R_{\mathrm{eq}}=0.85$.
	\item Under the assumption of $R_{\mathrm{eq}}=0.85$, the
          median Be star age drops from $261.7\;\mathrm{Myr}$ to
          $35.4\;\mathrm{Myr}$ (Fig. \ref{fig:obstime}). The drastic
          difference can be explained by the steepness of the
          mass-main sequence lifetime relation
          $\tau_{\mathrm{MS}}\propto M^{-2.5}$
          (e.g. \citealp{hansen_stellar_2004}).
	\item \citet{pols_formation_1991} point out that the
          majority of Be stars formed through the post-mass-transfer
          scenario are still bound to their companion, because the
          primary normally transfers sufficient mass to the secondary to either avoid the
          supernova or for the reduced mass loss in the supernova to not
          unbind the system. These companions are white dwarfs,
          neutron stars and black holes. The re-weighting changes both
          the fraction of the observed Be star population predicted to
          be in binaries and the stellar types of those companions
          (Fig. \ref{fig:obspartners}). Assuming
          $R_{\mathrm{eq}}=0.85$, our simulation predicts that more
          than 90\% of all Be stars are in binaries, but that among
          our observed sample only 77.1\% are in
          binaries. Furthermore, the observed binaries are less than
          half as likely to contain a Be star with a white dwarf
          companion. The companion that transfers mass to form the Be
          star must be low-mass in order for the remnant to be a white
          dwarf, thus the companion Be star must itself be low-mass
          and so late-type. The high fraction of Be stars residing in
          binaries with neutron stars or black holes agrees
          qualitatively with observations of high mass X-ray binaries
          (HMXBs) in the Magellanic Clouds. In fact, 33 of the 40
          HMXBs in the Large Magellanic Cloud
          \citep{antoniou_star_2016} and 69 of the 70 HMXBs in the
          Small Magellanic Cloud \citep{antoniou_star_2016} have Be
          star companions. Curiously, these authors note that the star
          formation efficiency of HMXBs in the Large Magellanic Cloud
          is 17 times smaller than in the Small Magellanic Cloud and
          argue that this is due to the difference in age and
          metallicity of the two stellar populations. The effect of
          varying metallicity on our predictions is an interesting
          avenue for future work.
\end{enumerate}

An important aspect of this work is that {\sc binary\_c} does not
carry out full stellar structure integration, instead opting for
prescriptions that permit the rapid evolution of a large number of
binary stars. This simplification allows for rapid population
  synthesis studies and places \textsc{binary\_c} in the class of
  synthetic binary stellar evolution codes (see Table 2 of
  \citealp{de_marco_dawes_2017} for a list of synthetic and detailed
  binary stellar evolution codes). Similar studies to the present one
  have been carried out with detailed binary stellar evolution codes,
  for instance by \citet{van_rensbergen_ob-runaways_1996} and more
  recently by \citet{eldridge_runaway_2011}.

\subsection{Compact object natal kick uncertainty}
\label{sec:uncertainty}

\begin{figure}
	\includegraphics[scale=0.55,trim = 4mm 4mm 0mm 3mm, clip]{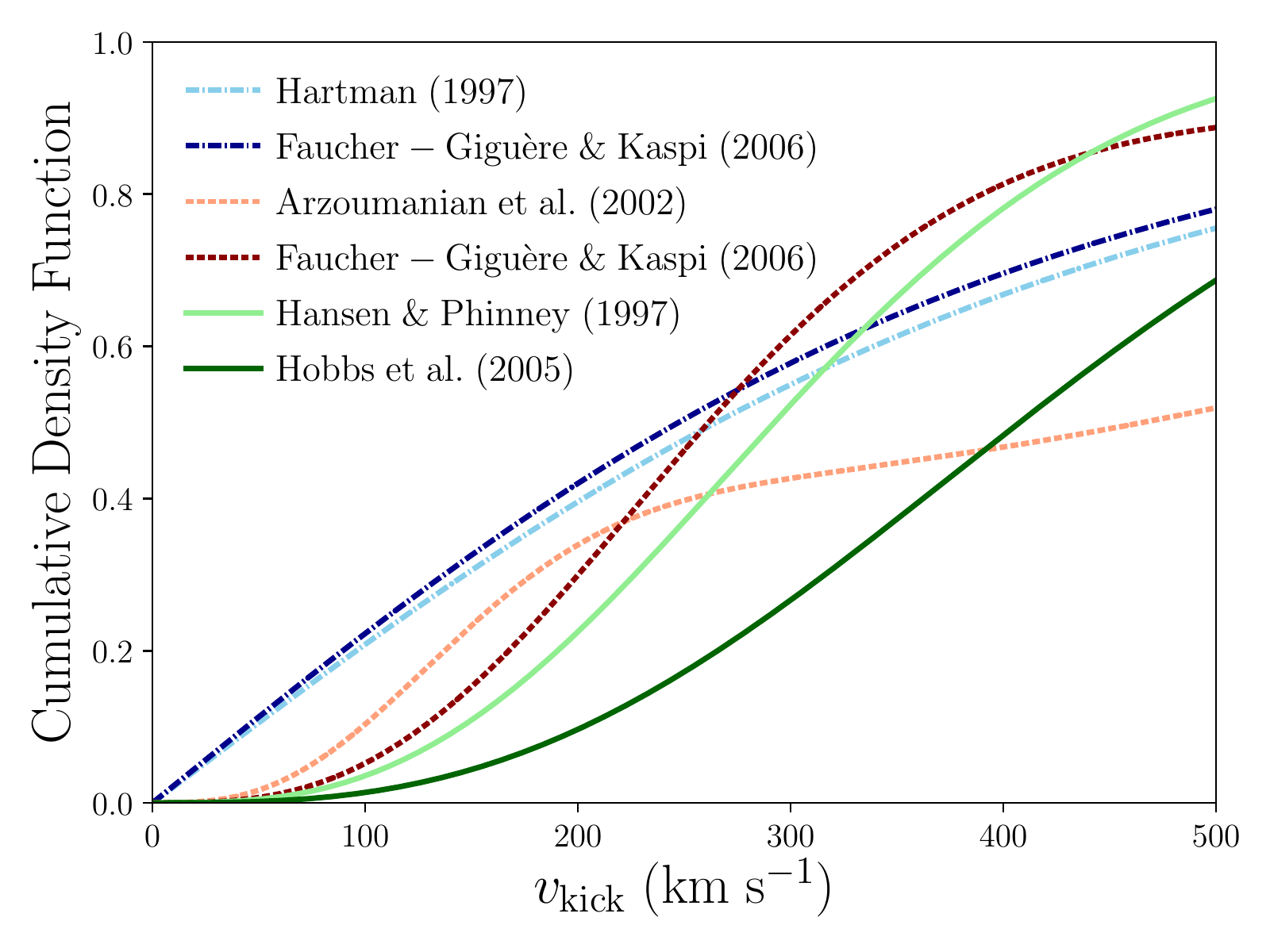}
	\caption{Various distributions for the 3D natal kick velocity
		distribution of neutron stars from the literature. The solid
		lines are the single Maxwellian distributions of
		\citet{hansen_pulsar_1997} and
		\citet{hobbs_statistical_2005}, the dash-dotted lines are from
		\citet{hartman_velocity_1997} and
		\citet{faucher-giguere_birth_2006}, and the dashed lines are
		the double Maxwellian distributions of
		\citet{arzoumanian_velocity_2002} and
		\citet{faucher-giguere_birth_2006}.}
	\label{fig:vejpulsar}
\end{figure}

There are several aspects of binary stellar physics that are
  poorly constrained observationally and must be prescribed when doing
  population synthesis. Three examples of relevance to the study of
  runaway stars are common envelope evolution, natal kicks of compact
  objects and fallback of material onto black holes. We attempt to
  quantify how different prescriptions alter our results by generating
  a population of binary stars with each prescription. However, common
  envelope evolution is sufficiently poorly understood that there are
  whole families of alternative prescriptions, so that an
  investigation of this uncertainty is beyond the scope of this
  paper. In contrast, for natal compact object kicks, we need only
  define the probability distribution of kicks
  $\operatorname{P}(v_{\mathrm{kick}})$.  There are several
  distributions in the literature that are fits of analytic models to
  data. We show a sample of the most popular distributions in
  Fig.~\ref{fig:vejpulsar}, including our fiducial choice of the
  Maxwellian distribution from \citet{hansen_pulsar_1997}. Clearly,
  the true natal kick distribution is not known at present -- even
  fits within the same work giving radically different distributions
  (i.e. \citealp{faucher-giguere_birth_2006})! A distribution with a
  large number of low-velocity compact objects predicts a high
  fraction of runaways with compact companions.
	
To explore the impact of the choice of distribution, we simulate a
  population of binary stars using the \citet{hartman_velocity_1997}
  distribution. This distribution has about 20\% of neutron stars
  receiving a kick less than $100\;\mathrm{km}\;\mathrm{s}^{-1}$ and
  thus we expect more binaries to survive the supernova of the
  primary. The most direct outcome of switching to the
  \citet{hartman_velocity_1997} prescription is that in the
  full model population, 0.6\% fewer Be stars are runaways
  and 0.6\% more are in binaries with neutron stars or black holes. In
  the population with selection effects included, this trend still holds true, with
  15.3\% Be star runaways and 17.6\% Be stars in compact object
  binaries. Another trend is a slight move in the runaway velocity
  distribution to slower velocities by roughly
  $0.5\;\mathrm{km}\;\mathrm{s}^{-1}$, which is a consequence of both
  the slower speeds of the neutron stars and the fact that energy is
  proportional to velocity squared. Even if two objects are unbound
  they still lose speed while escaping, and the speed lost is greater
  if the two objects have a smaller relative velocity.

Fallback of material onto a black hole in the seconds after the
  supernova can damp the birth kick and potentially lead to binaries
  surviving supernovae. One prescription for this fallback was
  presented in \citet{fryer_compact_2012}, who noted that it can alter
  the predicted number of $2-5\;\mathrm{M}_{\odot}$ black holes in
  X-ray binaries by tens of percent (Be stars are frequently found in
  X-ray binaries). We have not included this fallback in our fiducial
  model because the uncertainty inherent in the fallback prescription
  couples to the uncertainty in both the mass of the compact remnant
  after a supernova (i.e. whether the remnant is a neutron star or a
  black hole) and the natal kick distribution. Exploring the full
  space of possible prescriptions is beyond the scope of this paper.
  However, we believe that including fallback should cause a change in
  the Be star runaway fraction comparable to changing the natal kick
  distribution to that of \citet{hartman_velocity_1997}.

We conclude that the uncertainty in the distribution of natal
  kicks of compact objects does not greatly impact our results, aside
  from decreasing the expected Be star runaway fraction by a few
  percent. One reason for the small scale of the change is that
  whether a binary is disrupted by a supernova is principally
  determined by whether the primary loses more than half its mass
  \citep{blaauw_origin_1961}, and the kick on the compact object is
  only a second order effect. The uncertainty in the fallback of
  material onto a black hole remnant could couple with the natal kick
  uncertainty, however, and the Be star runaway fraction predicted for
  the observed catalogue could be as low as 10\%.

\section{Bayesian Framework}
\label{sec:bayesian}

We now ask what fraction of the observed Be stars are runaway
  stars, which is closely related to the question of what fraction can
  have their origins with the post-mass-transfer model. This is a
more subtle question than simply looking for Be stars with peculiar
velocities greater than $40\;\mathrm{km}\;\mathrm{s}^{-1}$, as done by
\citet{berger_search_2001}, because Fig.~\ref{fig:obsrunvel} shows
that the median runaway velocity of our simulated Be runaway
stars (accounting for observational biases) is only
$19.6\;\mathrm{km}\;\mathrm{s}^{-1}$. \citet{aumer_kinematics_2009}
used Hipparcos astrometry and photometry and Geneva-Copenhagen radial
velocities to calculate the velocity dispersion as a function of
colour of 15,000 nearby main-sequence and subgiant stars. For blue
stars $B-V<0$, they find that the total velocity dispersion is less
than $20\;\mathrm{km}\;\mathrm{s}^{-1}$. We would thus expect only a
small fraction of Be star runaways to have a peculiar velocity in
excess of $40\;\mathrm{km}\;\mathrm{s}^{-1}$. Another factor is that,
in the post-mass-transfer scenario, most Be star producing binaries
either evade the supernova through mass transfer or remain bound
post-supernova. Thus, most Be stars are found in binaries with white
dwarfs, neutron stars or black holes
(Fig. \ref{fig:obspartners}). This means that only a few percent of
all Be stars need have peculiar velocities in excess of
$40\;\mathrm{km}\;\mathrm{s}^{-1}$ in order to imply that all Be stars
are produced through the post-mass-transfer channel.

\subsection{Bayesian Mixture Model for the Kinematics}
\label{sec:bekinematics}

The null hypothesis for a young star in the Milky Way disc is that the
velocity is well described by the disk rotation plus velocities drawn
from a velocity ellipsoid centred on zero with dispersions
$(\sigma_{\mathrm{R}},\sigma_{\phi},\sigma_{\mathrm{z}})$.  The
velocity dispersions increase with the age on timescales of the order
of $1\;\mathrm{Gyr}$. We can reasonably neglect the age dependency and
thus assume the dispersions are constant.

We construct a model for a population of stars in the thin disc built
from three sub-populations. A fraction $F_{\mathrm{disc}}$ are
well-described by the null hypothesis, a fraction $F_{\mathrm{run}}$
are runaways and have an additional randomly oriented velocity
$v_{\mathrm{run}}$, and a fraction $F_{\mathrm{out}}$ are outliers and
have a different randomly oriented velocity
$v_{\mathrm{out}}$. One physical interpretation of any outliers is
  that they are stars that have under-gone the two-step-ejection
  scenario of \citet{pflamm-altenburg_two-step_2010}, in which a
  massive binary is ejected from a cluster by a dynamical interaction
  and the companion is later accelerated a second time by the
  supernova of the primary.  There are only two free parameters that
describe these fractions because they sum to unity and so form a
simplex.  We assume that every point in this simplex is equally
likely, which can be implemented as a flat Dirichlet prior. The
probability of observing a star with observables $\boldsymbol{x}$
under this composite model is simply a linear combination of the three
probability density functions,
\begin{equation}
P(\boldsymbol{x})=F_{\mathrm{disc}}P(\boldsymbol{x}\lvert
\mathrm{disc}) + F_{\mathrm{run}}P(\boldsymbol{x}\lvert
\mathrm{runaway})+ F_{\mathrm{out}}P(\boldsymbol{x}\lvert
\mathrm{outlier}).
\end{equation}
The motivation for including an outlier population is that in
Sec.~\ref{sec:threesources}, we removed three stars which had
obviously erroneous radial velocities
$|v_{\mathrm{rad}}|>500\;\mathrm{km}\;\mathrm{s}^{-1}$. However, some
spuriously large velocities may remain and boost the inferred runaway
fraction. The functions $P(\boldsymbol{x}\lvert\mathrm{model})$ are
the priors for the model parameters multiplied by the likelihood of the
data with those model parameters.

The observable properties of each star are the parallax $\omega$,
proper motions $(\mu_{\alpha\ast},\mu_{\delta})$ and radial velocity
$v_{\mathrm{rad}}$. The quantities we have are imperfect measurements
of the true parallax $\tilde{\omega}$, the true proper motions
$(\tilde{\mu}_{\alpha\ast},\tilde{\mu}_{\delta})$ and the true radial
velocity $\tilde{v}_{\mathrm{rad}}$. The true heliocentric distance
$\tilde{d}$ is then $1/\tilde{\omega}$ and, independently, the true
Galactocentric velocities
$(\tilde{v}_{\mathrm{R}},\tilde{v}_{\phi},\tilde{v}_{\mathrm{z}})$ can
be found using the equations in \citet{johnson_calculating_1987}. Both
these transforms are bijective, thus we are free to express our model
in either the distance or parallax and with either of the sets of the
velocities. Note that the bijectivity is conditional on the other
nuisance parameters having been fixed, such as the Solar position
$R_{\odot}$. Theoretically, the choice does not change the result of
the Bayesian inference and only switches our priors and
likelihoods. However, for practical reasons discussed later, it is
advantageous to pick the parameters which give the tightest prior. We
choose to express our prior in the parameters
$(\tilde{d},\tilde{\mu}_{\alpha\ast},\tilde{\mu}_{\delta},\tilde{v}_{\mathrm{rad}})$
and thus the likelihood in the parameters
$(\tilde{\omega},\tilde{v}_{\mathrm{R}},\tilde{v}_{\phi},\tilde{v}_{\mathrm{z}})$. We
include the necessary Jacobian $k^2$, where $k \approx 4.74057$ is the
conversion factor between $\mathrm{AU}\;\mathrm{yr}^{-1}$ and
$\mathrm{km}\;\mathrm{s}^{-1}$.

\subsection{The Priors}

Consider a model that has parameters $\boldsymbol{\theta}$ that makes
a prediction for some observable data $\boldsymbol{x}$. If we have a
function $\operatorname{Prior}(\boldsymbol{\theta})$ that describes
our prior expectation of the values that the parameters can take and a
function
$\operatorname{Likelihood}(\boldsymbol{x}\lvert\boldsymbol{\theta})$
that says how likely the data $\boldsymbol{x}$ is given a specific
choice of the parameters, then our posterior knowledge of the
parameters after measuring the data is specified through the
distribution
\begin{equation}
\operatorname{Posterior}(\boldsymbol{\theta}\lvert\boldsymbol{x})\propto \operatorname{Prior}(\boldsymbol{\theta})\operatorname{Likelihood}(\boldsymbol{x}\lvert\boldsymbol{\theta}).
\end{equation}
If new data are subsequently taken, then this posterior distribution
should be used as the prior when incorporating the new data. Through
this iterative process our knowledge of the parameters converges on
their true values.

The parameters can be split into global parameters
$\boldsymbol{\theta}^{\mathrm{g}}$ of the entire population and local
parameters $\boldsymbol{\theta}^{\mathrm{l}}$ of each star. The global
parameters only appear once in the prior. However, if the model is
applied to $N$ stars then $N$ independent copies of the local
parameters must be included and are labelled as
$\boldsymbol{\theta}^{i}$.

\subsubsection{Global Parameters} 

The peculiar velocity of the population at birth is assumed to be
Gaussian in each of the radial, azimuthal and vertical directions,
centred on zero and with independent dispersions
$(\sigma_{\mathrm{R}}, \sigma_{\phi}, \sigma_{\mathrm{z}})$. We place
a weakly informative Gaussian prior centred on
$10\;\mathrm{km}\;\mathrm{s}^{-1}$, with a
$10\;\mathrm{km}\;\mathrm{s}^{-1}$ dispersion and bounded below at
zero on each of these dispersions. An additional hyper-parameter is
the characteristic length-scale $L$ of the population, for which we
assume a weakly informative Gaussian prior centred on
$0.5\;\mathrm{kpc}$, with a $0.5\;\mathrm{kpc}$ dispersion and bounded
below at zero. \citet{astraatmadja_estimating_2016} found $L\approx1.35$ for a simulation of the contents of the full Gaia catalogue and thus the prior of $L$ was motivated by the lower magnitude limit of TGAS. There are five nuisance parameters whose priors are
Gaussians centred on their measured value and with a dispersion given
by the measurement error: the rotation of the Galactic disc
$V_{\mathrm{c}}$, the Galactocentric radius of the Sun $R_{\odot}$,
and the peculiar velocities of the Sun $(U_{\odot}, V_{\odot},
W_{\odot})$. We assume that the Milky Way disc rotates with a flat
circular velocity of
$V_{\mathrm{c}}=238\pm9\;\mathrm{km}\;\mathrm{s}^{-1}$ and that the
Sun orbits at the Galactocentric radius
$R_{\odot}=8.27\pm0.29\;\mathrm{kpc}$ with a peculiar velocity
$(U_{\odot}, V_{\odot},
W_{\odot})=(11.1\pm0.75\pm1,12.24\pm0.47\pm2,7.25\pm0.37\pm0.5)\;\mathrm{km}\;\mathrm{s}^{-1}$
\citep{schonrich_local_2010,schonrich_galactic_2012}.

\subsubsection{Local Parameters} 

For the true heliocentric distance $\tilde{d}$, we use the
exponentially decreasing volume density prior of
\citet{bailer-jones_estimating_2015}
\begin{equation}
P(\tilde{d}) = 
\begin{cases}
\frac{\displaystyle \tilde{d}^2}{\displaystyle 2L^3}\exp \left(-\tilde{d}/L \right), & \text{if}\ \tilde{d}>0,\\
\null & \null \\
0, & \text{otherwise},
\end{cases}
\end{equation}
where $L$ is the characteristic length scale hyper-parameter. We
remark that this can be equivalently stated in terms of the Gamma
distribution as $\mathrm{Gamma}(3,L)$. For the true proper motions and
radial velocity, the prior is a Gaussian centred on the measured value
and with a dispersion given by the measurement error. Both the runaway
and outlier populations have an additional velocity described by a
speed and a unit vector. The unit vector of ejection
$\boldsymbol{x}_{\mathrm{ej}}$ is assumed to be uniformly distributed
on the unit sphere. The runaway speed $v_{\mathrm{run}}$ is drawn from
the analytic distribution obtained in Sec.~\ref{sec:vejfit}. The
outlier speed $v_{\mathrm{out}}$ is drawn from a Gaussian centred on
zero and with a $500\;\mathrm{km}\;\mathrm{s}^{-1}$ dispersion. Both
speeds are constrained to be greater than zero.

In summary, the prior is
\begin{equation}
\operatorname{Prior}(\boldsymbol{\theta})=\operatorname{Global}(\boldsymbol{\theta}^{\mathrm{g}})\prod_i\operatorname{Local}(\boldsymbol{\theta}^{i}),
\end{equation}
where
\begin{align}
\operatorname{Global}(\boldsymbol{\theta}^{\mathrm{g}})=&\operatorname{Simplex}(F_{\mathrm{disc}},F_{\mathrm{run}},F_{\mathrm{out}})\operatorname{Normal}(\sigma_{\mathrm{R}},\sigma_{\phi},\sigma_{\mathrm{z}}) \nonumber\\
&\times\operatorname{Normal}(L)\operatorname{Normal}(V_{\mathrm{c}},R_{\odot},U_{\odot},V_{\odot},W_{\odot}),
\end{align}
and
\begin{align}
\operatorname{Local}(\boldsymbol{\theta}^{\mathrm{l}})=&\operatorname{Gamma}(\tilde{d})\operatorname{Normal}(\tilde{\mu}_{\alpha\ast},\tilde{\mu}_{\delta},\tilde{v}_{\mathrm{rad}}) \nonumber\\ &\times\operatorname{DLN}(v_{\mathrm{run}})\operatorname{Normal}(v_{\mathrm{out}})\operatorname{UnitVector}(\boldsymbol{x}_{\mathrm{ej}}).
\end{align}
All parameters except the fractions
$(F_{\mathrm{disc}},F_{\mathrm{run}},F_{\mathrm{out}})$ and the three
components of the unit vector $\boldsymbol{x}_{\mathrm{ej}}$ are
independent and the groupings have only been made to simplify the
notation. The function $\operatorname{DLN}(\cdot)$ indicates the
double log-normal fit to the numerical runaway velocity distribution
discussed in Sec.~\ref{sec:vejfit}.

\subsection{The Likelihood}

The likelihood gives the probability of observing the data
$\boldsymbol{x}$ given the model parameters
$\boldsymbol{\theta}$. Note that in this problem the data can be
broken down into a set of local data $\boldsymbol{x}^{\mathrm{l}}$ for
each star whose likelihood only depends on the global parameters
$\boldsymbol{\theta}^{\mathrm{g}}$ and the local parameters
$\boldsymbol{\theta}^{\mathrm{l}}$ of that star. Continuing the
notation above, we use $\boldsymbol{x}^{i}$ and
$\boldsymbol{\theta}^{i}$ to label the local data and parameters of a
specific star $i$. The likelihood is the product of the likelihoods of
each star with each individual likelihood containing four terms. The
first states how likely the observed parallax $\omega$ is given the
true parallax $\tilde{\omega}$ and is a Gaussian centred on the true
parallax with the measurement error on the parallax as the
dispersion. The remaining three terms give the likelihood of the
peculiar velocity and are Gaussians centred on zero and with
dispersions given by the parameters $(\sigma_{\mathrm{R}},
\sigma_{\phi}, \sigma_{\mathrm{z}})$. In summary the likelihood is,
\begin{equation}
\operatorname{Likelihood}(\boldsymbol{x}\lvert\boldsymbol{\theta})=\prod_i\operatorname{Likelihood}(\boldsymbol{x}^i\lvert\boldsymbol{\theta}^{\mathrm{g}},\boldsymbol{\theta}^{i}),
\end{equation}
where
\begin{align}
\operatorname{Likelihood}(\boldsymbol{x}^{\mathrm{l}}\lvert\boldsymbol{\theta}^{\mathrm{g}},\boldsymbol{\theta}^{\mathrm{l}})&=k^2\operatorname{Normal}(\omega)\left[F_{\mathrm{disc}}\operatorname{Normal}(\tilde{v}_{\mathrm{R}},\tilde{v}_{\phi},\tilde{v}_{\mathrm{z}})\right. \nonumber\\
&\left.+F_{\mathrm{run}}\operatorname{Normal}(\tilde{v}_{\mathrm{R}}-v_{\mathrm{run,R}},\tilde{v}_{\phi}-v_{\mathrm{run,}\phi},\tilde{v}_{\mathrm{z}}-v_{\mathrm{run,z}})\right. \nonumber\\
&\left.+F_{\mathrm{out}}\operatorname{Normal}(\tilde{v}_{\mathrm{R}}-v_{\mathrm{out,R}},\tilde{v}_{\phi}-v_{\mathrm{out,}\phi},\tilde{v}_{\mathrm{z}}-v_{\mathrm{out,z}})\right].
\end{align}
The groupings are only made to simplify the notation and the $k^2$ is
the Jacobian of the transformation.

\subsection{Analytic Fit to Runaway Velocity Distributions}
\label{sec:vejfit}

The final ingredient that we need before application of Hamiltonian
Monte Carlo techniques is an analytic fit to the numerical
distribution of simulated runaway velocities from Sec.~\ref{sec:binaryc}. We specifically refer to the runaway velocity distribution for the model Be star population defined by $R_{\mathrm{eq}}=0.85$ accounting for the observational selection effect, after averaging across the different stellar types. This
step is necessary as the derivatives of the posterior with respect to
the parameters need to be calculable.

\begin{figure}
	\includegraphics[scale=0.55,trim = 4mm 5mm 0mm 3mm,
          clip]{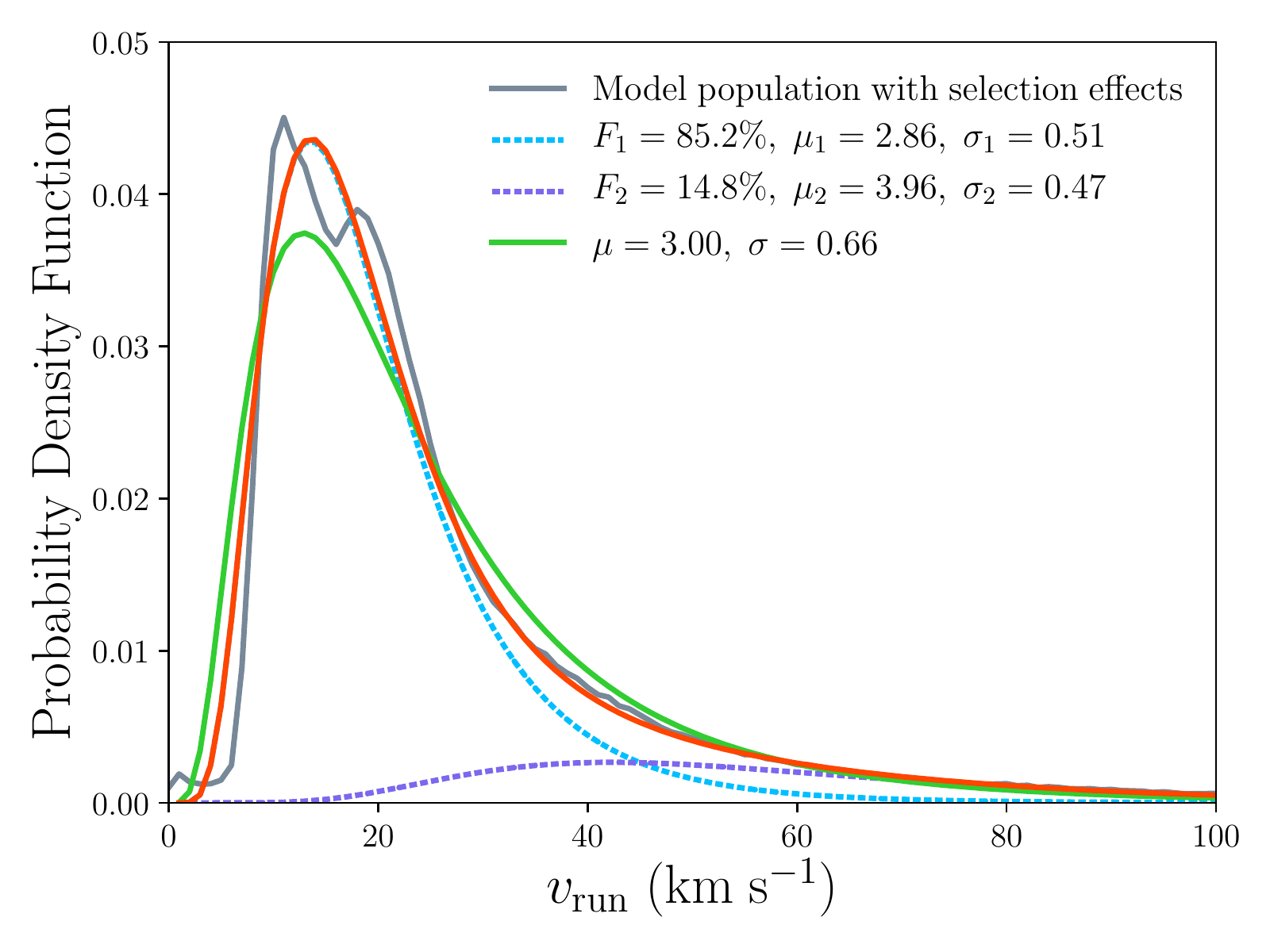}
	\caption{Fits of one and two log-normal distributions to the
          numerical predicted runaway velocity distribution calculated
          in Section \ref{sec:binaryc}. In the legend, the $\mu$ and
          $\sigma$ give the parameters of each log-normal distribution
          and $F_1$ and $F_2$ are the fraction of the probability
          assigned to each component of the double log-normal fit.}
	\label{fig:vejfit}
\end{figure}

We opt to use the log-normal distribution, defined by the probability
density function,
\begin{equation}
\mathrm{P}(x\;|\;\mu,\sigma)=
\begin{cases}
\frac {\displaystyle 1 }{\displaystyle x\sigma\sqrt{2\pi}}\exp \left(-\frac{\displaystyle \left(\ln x-\mu\right)^2}{\displaystyle 2\sigma^2}\right), & \mathrm{if}\; x>0,\\
\null & \null \\
0, & \mathrm{otherwise},
\end{cases}
\end{equation}
and with cumulative density function,
\begin{equation}
\mathrm{F}(x\;|\;\mu,\sigma)=
\begin{cases}
\frac12 + \frac12\operatorname{erf}\Big[\frac{\displaystyle \ln x-\mu}{\displaystyle \sqrt{2}\sigma}\Big], &\mathrm{if}\;x>0,\\
0, & \mathrm{otherwise}.
\end{cases}
\end{equation}
The log-normal distribution has several properties which make it
appropriate for modelling a runaway velocity distribution:
\begin{enumerate}
	\item log-normals are constrained to be non-zero only for
          positive $x$,
	\item the mode is given by $$\exp{\left(\mu-\sigma^2\right)}$$
          and thus for $\mu<<\sigma^2$ a log-normal can approximate a
          decay,
	\item the skewness is given
          by $$\left(\exp{\sigma^2}+2\right)\sqrt{\vphantom{\sum}\exp{\sigma^2}-1}$$
          and so for small $\sigma$ a log-normal can be used to
          approximate a symmetric distribution offset from the origin.
\end{enumerate}
We fit a mixture of two log-normal distributions to the numeric
runaway velocity distribution using the implementation of non-linear
least squares in {\sc SciPy}. We found that the fit was best performed
using the cumulative density function. In Fig.~\ref{fig:vejfit} we
show a comparison between the probability density functions of our
numeric distribution compared to the sum of the two log-normal
distributions, as well as a fit containing a single log-normal
distribution to illustrate the improvement. The double log-normal
  does not capture the bimodality near the peak of the distribution,
  but does trace the tail to large velocities.

\section{The Fraction of Be Runaways}
\label{sec:results}

\subsection{Preliminaries}

363 of the 632 stars in our observed sample have TGAS astrometry
and thus have a published covariance matrix between their position,
parallax and proper motions. We neglect the covariance matrix here,
but plan to revisit the problem after the second Gaia data release,
which will include the vast majority of the known Be stars. Positions
in TGAS are measured with $\mathrm{milliarcsecond}$ accuracy which at
a distance of $1\;\mathrm{kpc}$ corresponds to a spatial accuracy of
$5\;\mu \mathrm{pc}$. The distance over which the properties of the
Galaxy vary is much larger than this and thus we fix the position of
each star.

Note that, in principle, we could condition the runaway velocity
  distribution for individual Be stars on their stellar type. However,
  there are two main arguments against this. First, the individual
  runaway velocity distributions are subject to greater Poisson noise
  than their weighted mean. Second, the spectral types for the Be
  stars in our sample were queried from SIMBAD and thus have a variety
  of provenances. Taken as a whole, they likely give a good summary of
  the spectral type distribution of the sample. However, each
  individual spectral type has a different and unknown uncertainty. If
  the sample had measured effective temperatures and luminosities with
  uncertainties, then it would be sensible to condition the runaway
  velocity distribution of individual Be stars on these physical
  properties. Such an approach will become viable with the second Gaia
  data release.

When our model is applied to the 632 stars in our observed Be
star database, we have 5067 free parameters. These are broken down
into eleven global parameters which describe the entire sample and
eight local parameters for each star. We used the Bayesian inference
platform {\sc Stan} \citep{carpenter_stan_2017}, accessed through {\sc
  CmdStan}\footnote{Stan Development Team. 2017. CmdStan: the
  command-line interface to Stan, Version
  2.16.0. \url{http://mc-stan.org}}, to obtain the posterior for the
model. {\sc Stan} includes a Hamiltonian Markov chain Monte Carlo
sampler which incorporates No-U-Turn Sampling
\citep{hoffman_no-u-turn_2011} and so is well suited to
high-dimensional problems. We ran four chains and computed the
potential scale reduction statistic $\hat{R}$
\citep{gelman_inference_1992} across the chains to assess
convergence. We used an equal number of warm-up and sampling
iterations and doubled the number of iterations until the model
converged. The four chains were then merged and used to calculate the
statistics.

\subsection{Retrieval of simulated data}
\label{sec:retrieval}
A minimum requirement of a successful Bayesian model is the
  ability to retrieve the input model parameters of simulated data.
  This is equivalent to saying that if our model of Be star
  kinematics is correct and complete, then the parameters we retrieve
  should be close to the true values. One common reason for this not
  to be true is the existence of a degeneracy between two parameters
  which the data are not precise enough to break. For our Be star
  model, we expect there to be a strong degeneracy between the three
  velocity dispersions and the runaway fraction, and thus it is
  essential that we investigate whether the uncertainties on our data
  are sufficiently small to allow this degeneracy to be broken.

We generate a test set of stars assuming known
  $F_{\mathrm{run}}=17.5\%$ and
  $(v_{\mathrm{R}},v_{\phi},v_{\mathrm{z}})=(12,10,4)\;\mathrm{km}\;\mathrm{s}^{-1}$. We
  draw a random set of
  $(V_{\mathrm{c}},R_{\odot},U_{\odot},V_{\odot},W_{\odot})$ from
  within their uncertainties to act as the `truth' of the fake
  catalogue. We create the fake stars in a one-to-one correspondence
  with the real catalogue, taking both the positions and uncertainties
  from the true stars. The size of the uncertainties roughly
  correlates with the distance to the star, so to sample a realistic
  distance to each fake star given the uncertainties, we draw from the
  posterior for the distance to the real star
  $\operatorname{Gamma}(0,L) \operatorname{Normal}
  (\omega,\sigma_{\omega})$, where we fix $L=0.2\;\mathrm{kpc}$ and
  $(\omega,\sigma_{\omega})$ refer to the parallax and uncertainty of
  the real star. To generate the radial velocities and proper motions,
  we first draw three random velocities from the velocity dispersion
  distributions, add on a randomly oriented ejection velocity for the
  randomly selected subset of runaway stars, and then transform to the
  equatorial frame. Finally, the parallaxes, radial velocities and
  proper motions are convolved with the uncertainties. Note that we do
  not include an outlier population.

\begin{figure*}
	\includegraphics[scale=0.7,trim = 0mm 0mm 0mm 0mm, clip]{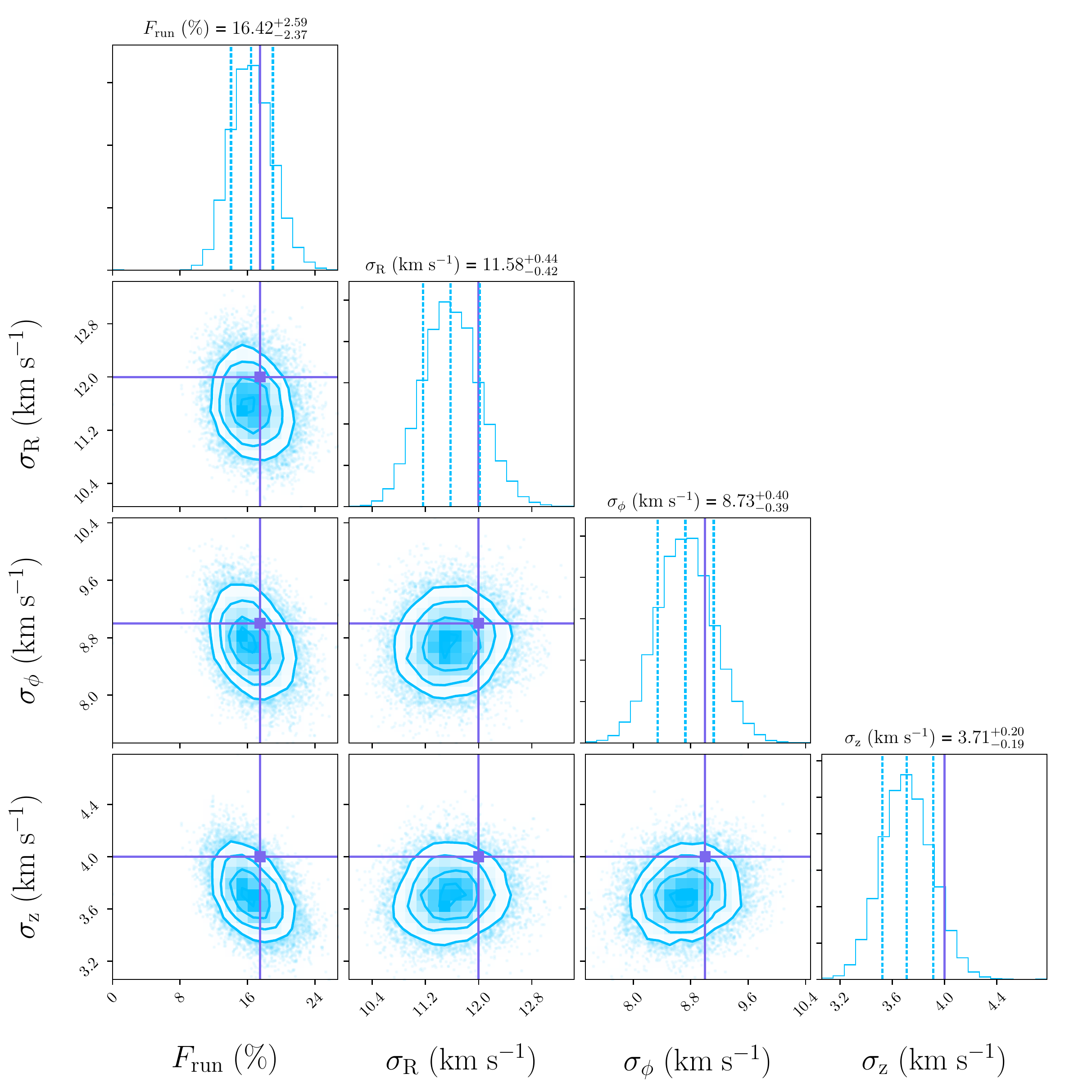}
	\caption{Corner plot of the posterior for the retrieval.}
	\label{fig:retrievecorner}
\end{figure*}

Our principal model is the one introduced in
  Sec.~\ref{sec:bayesian}, which has contributions from disk, outlier
  and runaway populations and which uses the double log-normal fit to
  the runaway velocity distribution. We present the posterior for this
  model when applied to the fake catalogue in
  Fig.~\ref{fig:retrievecorner}, with the true values shown in
  purple. The interpretation of this plot is that the uncertainties on
  the current set of Be stars are too large to wholly break the
  degeneracy between $F_{\mathrm{run}}$ and the velocity
  dispersions. However, given the width of the priors on these
  parameters, the retrieved values are remarkably in agreement with
  the true values.

\begin{figure}
	\includegraphics[scale=0.78,trim = 22mm 0mm 0mm 10mm,
	clip]{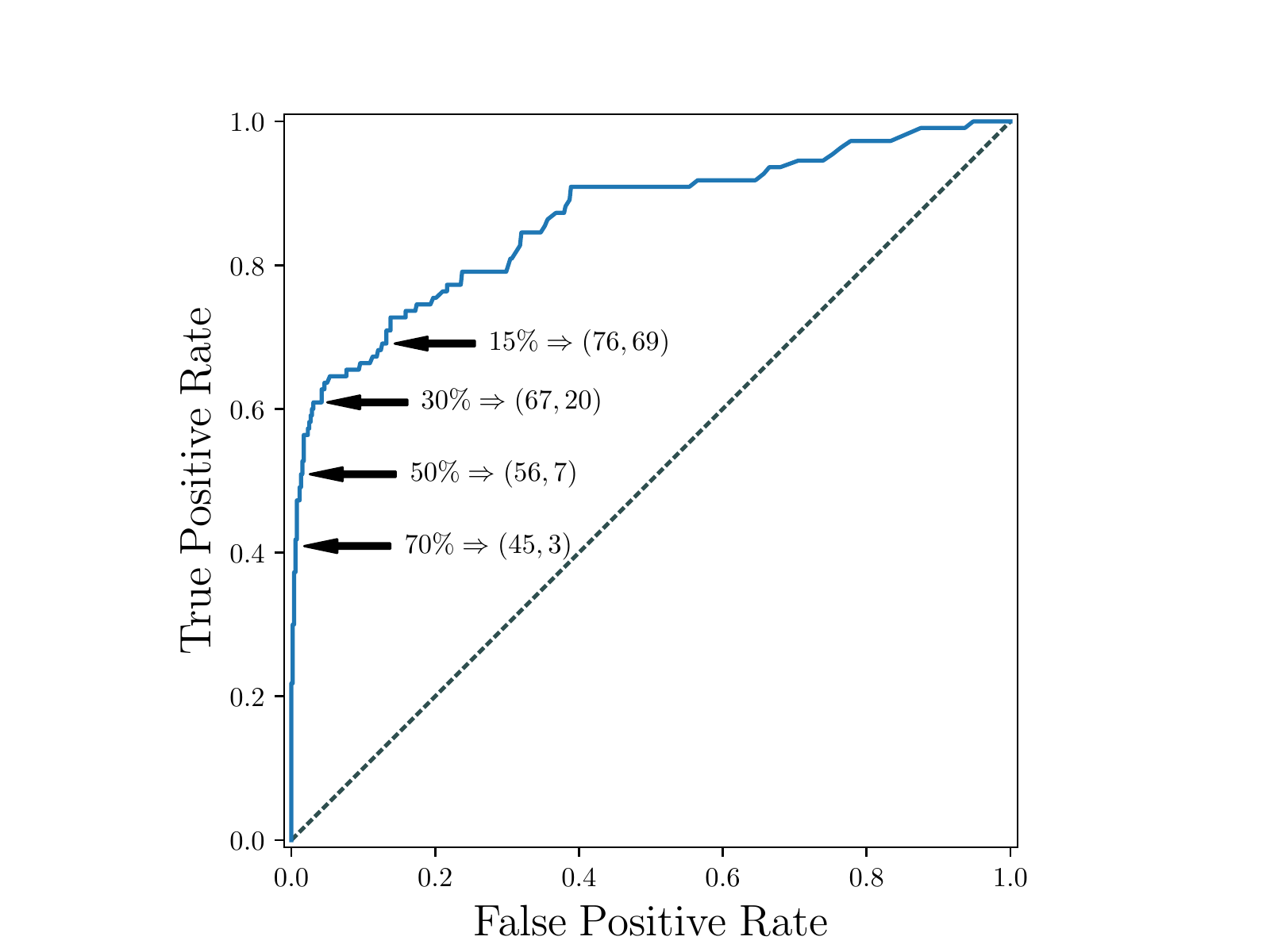}
	\caption{The receiver operating characteristic curve for the
          classification of runaway stars based on the mean of the
          posterior probability of being a runaway star being greater
          than $x\%$. The grey dashed line corresponds to a classifier
          that randomly guesses and which our classification far
          out-competes. The annotations are of the form
          `$x\%\Rightarrow(\#\;\mathrm{of}\;\mathrm{True}\;\mathrm{Positives},\#\;\mathrm{of}\;\mathrm{False}\;\mathrm{Positives})$'.}
	\label{fig:roccurve}
\end{figure}

While our chief concern is the runaway fraction among the entire
  Be star population, it is interesting to consider whether our method
  is successful at classifying individual stars. We expect that stars
  are only successfully classified as runaways if they are travelling
  sufficiently rapidly, but not so rapidly as to be misclassified as
  outliers. Our method does not classify stars directly, but does
  return the probability of membership of each class. An obvious
  classification scheme is that if the probability of being a runaway
  star is greater than $x\%$, then the star is classed as a
  runaway. One common metric to decide what value of $x$ to use is the
  receiver operating characteristic curve (Fig. \ref{fig:roccurve}),
  where the true positive rate and false positive rate are plotted as
  a parametric function of $x$. A classification scheme where any star
  with a probability of being a runaway star greater than 50\% is
  classed as a runaway star performs well, because it can pick out
  more than half the true runaway stars with only 10\% contamination.

We conclude both that our method can accurately retrieve the
  fraction of runaway stars and the velocity dispersions, as well as
  produce a low-contamination list of runaway stars.

\subsection{Principal model}
\label{sec:principal}

 \begin{figure*}
	\includegraphics[scale=0.7,trim = 0mm 0mm 0mm 0mm, clip]{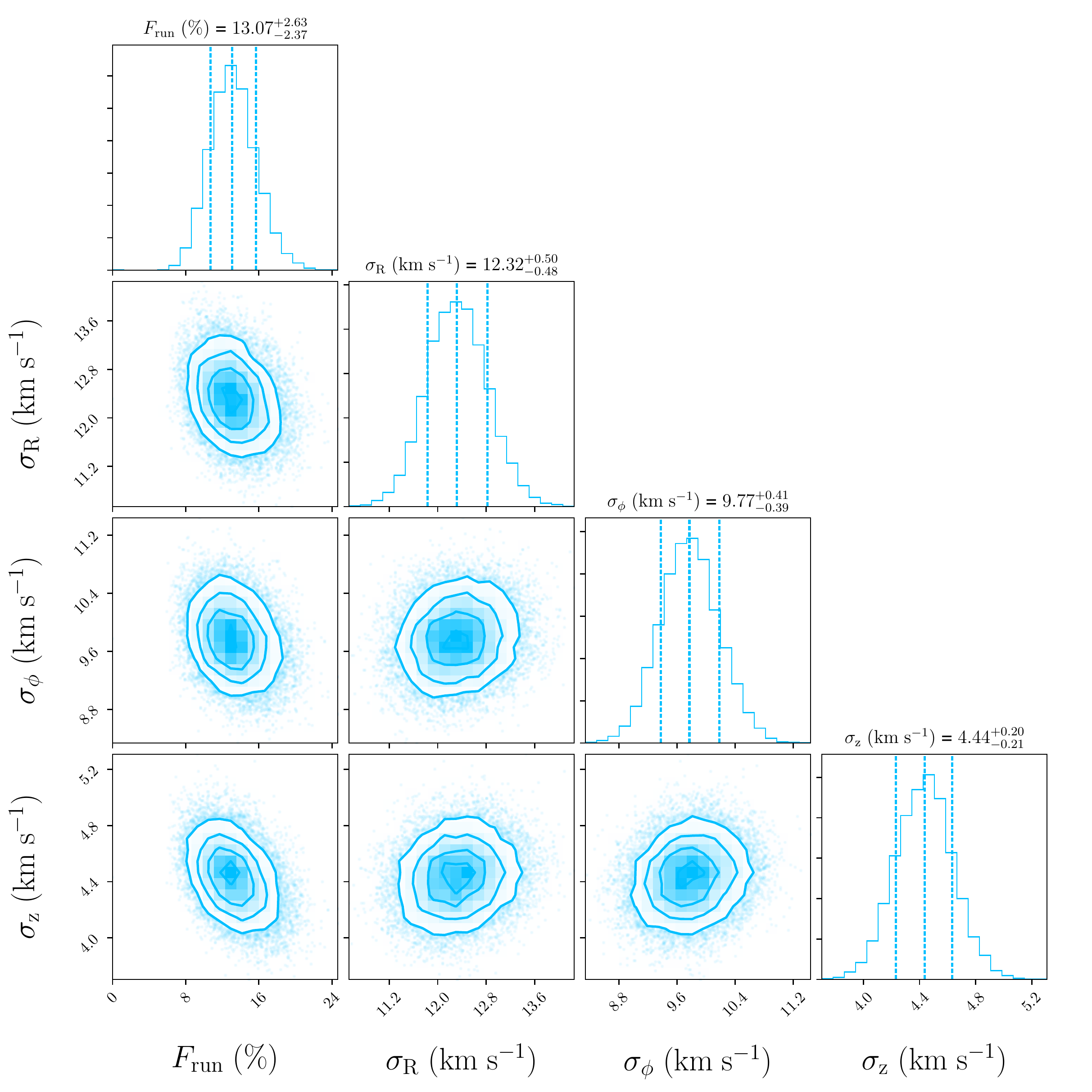}
	\caption{Corner plot of the posterior for the principal
          model.}
	\label{fig:corner}
\end{figure*}

We now apply the model to the true catalogue of Be stars described in
Section \ref{sec:data}. In Fig.~\ref{fig:corner}, we show the
posterior for the runaway fraction $F_{\mathrm{run}}$ and the velocity
dispersions of the population. The outlier fraction is only
$F_{\mathrm{out}}=0.36\substack{+0.35 \\ -0.22}\%$ which means that
the high velocity stars are sufficiently well described by the runaway
model. The posterior runaway fraction is about
$13.1^{+2.6}_{-2.4}\%$. Using the classification scheme from
Sec. \ref{sec:retrieval}, 40 stars are classified as runaway stars and
they are given in Table \ref{tab:runaway}. 7 of these stars were
classed as high peculiar space velocity stars by
\citet{berger_search_2001} and this subset are indicated in Table
\ref{tab:runaway}. We perform a brief search of SIMBAD to look for
particularly interesting cases:
\begin{itemize}
	\item \textbf{Menkib:} There are only 313
          IAU-approved\footnote{\url{https://www.iau.org/public/themes/naming_stars/},
            accessed 12/03/2018.} named stars and thus the appearance
          of one in a modern astronomical paper is remarkable. Menkib
          is a previously known runaway star
          \citep{hoogerwerf_origin_2001} in the constellation of
          Perseus and associated with the $2.5^{\circ}$ bowshock
          nebula NGC 1499 (otherwise known as the California Nebula).
	\item \textbf{Eclipsing binaries:} W Del is a member of an
          Algol eclipsing binary, while V716 Cen and RY Sct are
          members of Beta Lyrae eclipsing binaries. Both these classes
          of eclipsing, semi-detached binaries contain one main
          sequence and one giant star, in which the giant is transferring
          mass onto the main sequence star. This mass transfer could
          plausibly spin up the companion and produce a Be star, which
          explains the occurrence of three of these rare binaries in
          the list. That these three systems are binaries does not
          necessarily imply they are not runaway stars; semi-detached
          binaries are necessarily close and thus it is plausible for
          these systems to have originated in a triple system which
          was unbound in the supernova of the third, most-massive
          star. Mass transfer from the third star onto these binaries
          may indeed have lead to these systems being so close by
          causing an in-spiral due to gas drag. Runaway binaries from
          massive triple systems would have modest runaway velocities
          because there is necessarily a limit to how close they can
          be to the primary. Conversely, it is also possible that
          these binaries could have been ejected from the core of a
          young cluster through dynamical interactions. Dynamical
          ejection has been suggested as the more likely source of
          binary runaway stars, although the triple supernova
          mechanism should produce some number of runaway binaries
          \citep{perets_runaway_2009}.
	\item \textbf{Z Her:} \citet{harmanec_shell_1972} conclude
          that this system is composed of a $5.4\;\mathrm{M}_{\odot}$
          Be star with a $0.66\;\mathrm{M}_{\odot}$
          companion. However, \citet{popper_orbits_1988} conclude that
          this is instead an AM Canus Venaticorum variable with masses
          $1.61+1.31\;\mathrm{M}_{\odot}$, which is the classification
          reported by SIMBAD.
\end{itemize}

One aspect of runaway stars kinematics not included in our model is
that runaway B stars are found at greater altitudes above the disc
than a typical B star. Reassuringly, the 40 most probable runaway
stars have a greater spread in altitudes than the rest of the sample
(standard deviations $0.23\;\mathrm{kpc}$ and
$0.10\;\mathrm{kpc}$). Aside from the resulting slightly broader
distribution of Galactic latitudes, the 40 runaway stars are otherwise
distributed similarly across the sky to the rest of the catalogue.

\begin{table}
	\centering
	\caption{List of stars identified as highly-likely runaway
          stars by the method described in
          Sec. \ref{sec:principal}. $P_{\mathrm{run}}$ is the
          probability that the star is a runaway, $v_{\mathrm{run}}$
          is the posterior runaway velocity and $d$ is the posterior
          distance inferred from the parallax. All of the names are
          resolvable by Simbad. The starred entries are high-peculiar
          space velocity stars mentioned by
          \citet{berger_search_2001}.  Based on
          Fig. \ref{fig:roccurve}, we should expect around 4 of these
          stars to be false positives.}
	\label{tab:runaway}
	\setlength\tabcolsep{5 pt}
	\begin{tabular}{llcrr}
		\hline
		$P_{\mathrm{run}} (\%)$ & Name               & $v_{\mathrm{run}}\;(\mathrm{km}\;\mathrm{s}^{-1})$ & $d\;(\mathrm{pc})$            & $z\;(\mathrm{pc})$            \\ \hline
		99.8                    & CD-30 850          & $28.0\substack{+7.5 \\ -6.2}$                      & $704\substack{+294 \\ -206}$  & $-660\substack{+193 \\ -276}$ \\
		99.3                    & W Del              & $32.5\substack{+10.2 \\ -9.2}$                     & $835\substack{+191 \\ -141}$  & $-196\substack{+33 \\ -45}$   \\
		99.1                    & HD 20340$^{\ast}$  & $36.8\substack{+10.7 \\ -8.1}$                     & $408\substack{+149 \\ -97}$   & $-335\substack{+80 \\ -122}$  \\
		99.0                    & HD 30677           & $25.3\substack{+7.4 \\ -6.5}$                      & $1005\substack{+277 \\ -206}$ & $-380\substack{+78 \\ -105}$  \\
		98.9                    & HD 195407$^{\ast}$ & $59.1\substack{+11.6 \\ -12.3}$                    & $1063\substack{+261 \\ -182}$ & $-23\substack{+4 \\ -6}$      \\
		98.8                    & HD 127617          & $40.2\substack{+16.3 \\ -12.4}$                    & $703\substack{+231 \\ -170}$  & $640\substack{+210 \\ -155}$  \\
		98.7                    & HD 137387          & $56.9\substack{+12.1 \\ -11.9}$                    & $384\substack{+41 \\ -35}$    & $-93\substack{+8 \\ -10}$     \\
		98.4                    & HD 57682           & $48.4\substack{+13.2 \\ -10.5}$                    & $551\substack{+170 \\ -113}$  & $25\substack{+8 \\ -5}$       \\
		98.1                    & Menkib             & $54.9\substack{+12.8 \\ -13.5}$                    & $411\substack{+101 \\ -70}$   & $-93\substack{+16 \\ -23}$    \\
		98.0                    & HD 216044          & $50.6\substack{+13.0 \\ -14.4}$                    & $1428\substack{+412 \\ -320}$ & $-91\substack{+20 \\ -26}$    \\
		97.7                    & HD 194057          & $58.1\substack{+13.0 \\ -13.6}$                    & $1616\substack{+412 \\ -304}$ & $128\substack{+33 \\ -24}$    \\
		96.7                    & V2123 Cyg          & $76.4\substack{+12.7 \\ -12.7}$                    & $1034\substack{+250 \\ -183}$ & $-89\substack{+16 \\ -21}$    \\
		95.7                    & HD 107348$^{\ast}$ & $32.1\substack{+13.8 \\ -11.5}$                    & $114\substack{+11 \\ -9}$     & $74\substack{+7 \\ -6}$       \\
		94.6                    & HD 205618$^{\ast}$ & $31.6\substack{+16.0 \\ -11.5}$                    & $955\substack{+242 \\ -181}$  & $-269\substack{+51 \\ -68}$   \\
		94.2                    & HD 181409          & $77.0\substack{+20.7 \\ -20.8}$                    & $569\substack{+114 \\ -86}$   & $92\substack{+18 \\ -14}$     \\
		93.2                    & V716 Cen           & $49.5\substack{+17.1 \\ -19.5}$                    & $267\substack{+35 \\ -28}$    & $30\substack{+4 \\ -3}$       \\
		93.1                    & HD 81753           & $27.6\substack{+12.6 \\ -9.2}$                     & $447\substack{+131 \\ -92}$   & $120\substack{+35 \\ -25}$    \\
		92.7                    & HD 150288          & $78.2\substack{+20.3 \\ -24.9}$                    & $685\substack{+296 \\ -193}$  & $-6\substack{+2 \\ -3}$       \\
		92.0                    & TYC 3146-824-1     & $28.1\substack{+13.2 \\ -9.8}$                     & $836\substack{+205 \\ -145}$  & $186\substack{+46 \\ -32}$    \\
		90.8                    & HD 210129$^{\ast}$ & $88.3\substack{+15.5 \\ -17.6}$                    & $211\substack{+32 \\ -25}$    & $-96\substack{+11 \\ -15}$    \\
		89.0                    & BD+22 3833         & $20.4\substack{+7.3 \\ -5.9}$                      & $186\substack{+28 \\ -22}$    & $-7\substack{+1 \\ -1}$       \\
		88.5                    & HD 50658$^{\ast}$  & $31.8\substack{+14.6 \\ -12.4}$                    & $273\substack{+64 \\ -43}$    & $94\substack{+22 \\ -15}$     \\
		87.3                    & BD-1 3834          & $23.3\substack{+11.4 \\ -8.3}$                     & $1101\substack{+282 \\ -210}$ & $-255\substack{+49 \\ -65}$   \\
		85.7                    & TYC 3327-2315-1    & $18.9\substack{+7.4 \\ -5.5}$                      & $190\substack{+15 \\ -13}$    & $-18\substack{+1 \\ -1}$      \\
		82.9                    & HD 305560          & $29.0\substack{+11.9 \\ -11.5}$                    & $1627\substack{+375 \\ -289}$ & $-38\substack{+7 \\ -9}$      \\
		79.4                    & 7 Vul              & $25.9\substack{+11.9 \\ -10.0}$                    & $347\substack{+67 \\ -48}$    & $7\substack{+1 \\ -1}$        \\
		79.4                    & BD+23 3183         & $26.7\substack{+17.3 \\ -11.0}$                    & $778\substack{+200 \\ -141}$  & $317\substack{+81 \\ -57}$    \\
		78.8                    & CD-42 11983        & $20.1\substack{+8.5 \\ -6.5}$                      & $536\substack{+266 \\ -170}$  & $-22\substack{+7 \\ -11}$     \\
		68.9                    & HD 175511          & $19.1\substack{+9.3 \\ -6.4}$                      & $402\substack{+43 \\ -36}$    & $158\substack{+17 \\ -14}$    \\
		67.0                    & HD 152505          & $21.8\substack{+14.1 \\ -8.4}$                     & $763\substack{+178 \\ -121}$  & $-48\substack{+8 \\ -11}$     \\
		64.6                    & RY Sct             & $130.1\substack{+18.6 \\ -111.8}$                  & $1361\substack{+387 \\ -276}$ & $-3\substack{+1 \\ -1}$       \\
		61.4                    & GW Vel             & $24.3\substack{+14.0 \\ -10.2}$                    & $1087\substack{+273 \\ -199}$ & $-17\substack{+3 \\ -4}$      \\
		61.4                    & CD-29 5159         & $26.1\substack{+15.7 \\ -11.7}$                    & $1911\substack{+660 \\ -468}$ & $-21\substack{+5 \\ -7}$      \\
		60.8                    & Z Her              & $24.7\substack{+14.5 \\ -10.6}$                    & $84\substack{+2 \\ -2}$       & $42\substack{+1 \\ -1}$       \\
		59.8                    & HD 37657$^{\ast}$  & $26.6\substack{+17.6 \\ -12.6}$                    & $698\substack{+163 \\ -114}$  & $83\substack{+19 \\ -13}$     \\
		56.9                    & BD-20 6251         & $18.3\substack{+10.1 \\ -6.2}$                     & $250\substack{+80 \\ -50}$    & $-177\substack{+36 \\ -57}$   \\
		56.7                    & PZ Gem             & $23.9\substack{+15.7 \\ -10.5}$                    & $773\substack{+253 \\ -172}$  & $21\substack{+7 \\ -5}$       \\
		51.9                    & V442 And           & $24.4\substack{+17.1 \\ -11.1}$                    & $1056\substack{+280 \\ -243}$ & $-276\substack{+64 \\ -73}$   \\
		51.4                    & BD+52 2280         & $18.9\substack{+11.1 \\ -6.9}$                     & $322\substack{+68 \\ -44}$    & $121\substack{+26 \\ -17}$    \\
		50.4                    & TYC 870-115-1      & $24.3\substack{+19.1 \\ -11.0}$                    & $304\substack{+54 \\ -39}$    & $287\substack{+51 \\ -37}$   \\ \hline
	\end{tabular}
\end{table}


Suppose we now change the principal model by setting the runaway
fraction to zero.  Then we obtain a significant outlier fraction of
$F_{\mathrm{out}}=2.85\substack{+0.78 \\ -0.66}\%$, implying that
around 18 stars of the 632 Be stars in the dataset would need an
alternative explanation for their velocity. In this case, the
posterior also favours somewhat higher velocity dispersions, closer to
the values for a $250\;\mathrm{Myr}$ old population of stars. Given
that the Be stars in our simulated binary evolution have a median age
of $35.4\;\mathrm{Myr}$, this seems unlikely.



Another simple change to the principal model is to replace the runaway
velocity distribution with the single log-normal fit from
Sec.~\ref{sec:vejfit}, and for this modified model we obtain
\begin{align}
F_{\mathrm{run}}&=12.9_{-2.3}^{+2.6}\%, \nonumber \\ (\sigma_{\mathrm{R}},\sigma_{\phi},\sigma_{\mathrm{z}})&=(12.3,9.8,4.5)\pm(0.5,0.4,0.2)\;\mathrm{km}\;\mathrm{s}^{-1}.
\end{align}
The results are entirely consistent with the double log-normal case, which is
perhaps not surprising since the PDFs for both distributions are
within a factor of two at almost all velocities (see
Fig. \ref{fig:vejfit}). The motivation for mentioning this possibility
is that a single log-normal is easier to work with both analytically
and computationally, and that it provides a benchmark for the
subsequent section.

\subsection{A Free Log-normal}

 \begin{figure*}
	\includegraphics[scale=1.65,trim = 0mm 0mm 0mm 0mm, clip]{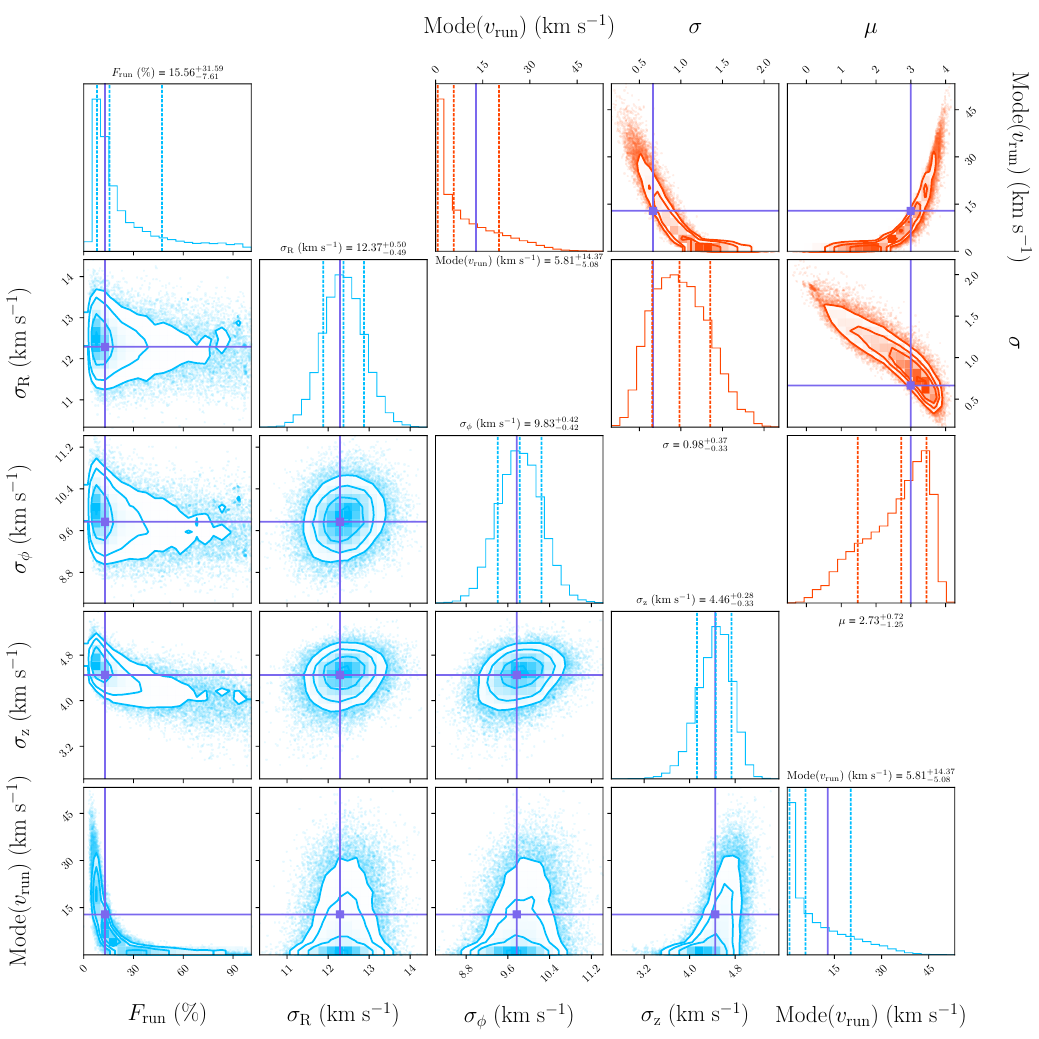}
	\caption{The posterior for a model in which the functional
          form of the runaway velocity distribution is constrained to
          be a $\operatorname{LogNormal}(\mu,\sigma)$ distribution,
          but the $\mu$ and $\sigma$ parameters are free to
          vary. \textbf{Upper:} The posterior distributions for $\mu$
          and $\sigma$ together with the resulting
          $\operatorname{Mode}(v_{\mathrm{run}})$. The green truths
          indicate the fixed values obtained by fitting a log-normal
          distribution to the simulated outcomes of binary
          evolution. \textbf{Lower:} The posterior for the runaway
          fraction and velocity dispersions with the mode of the
          runaway velocity distribution shown for reference with the
          upper panel. The green truths indicate the median values
          obtained from the model where $\mu$ and $\sigma$ were fixed
          as in Sec.~\ref{sec:vejfit}.}
	\label{fig:invertedfree}
\end{figure*}

The runaway velocity distribution obtained through binary evolution in
Sec.~\ref{sec:binaryc} has systematic errors due to gaps in our
understanding of the physics of interacting, massive binary stars as
well as the ad hoc definition of a Be star. This uncertainty can be
quantified by loosening the requirement that the runaway velocity
distribution is precisely the double log-normal obtained by our
fit. Instead, we assume that the runaway velocity distribution is
described by a log-normal distribution with two hyperparameters $\mu$
and $\sigma$. These hyperparameters have priors
$\mu\sim\operatorname{Normal}(0,10)$ and
$\sigma\sim\operatorname{Normal}(0.5,0.5)$. However, the data are
sufficiently informative that the choice of these priors is not
dominant. The posterior for this model is shown in the two corner
plots of Fig.~\ref{fig:invertedfree}.

The mode of the posterior in $(\mu,\sigma)$ is close to the values
which would replicate the single log-normal fit shown in
Fig.~\ref{fig:vejfit}. This implies that the runaway velocity
distribution predicted in Sec.~\ref{sec:binaryc} is plausible. One
of the key degeneracies in the posterior is between the runaway
fraction and the mode of the runaway velocity distribution. A property
of the log-normal distribution is that for $\mu<<\sigma^2$ the PDF can
approximate a decay from the origin, i.e. all the probability
is in runaway stars which have a zero runaway velocity, in which case
all the Be stars can be runaway stars with no increase in their
expected velocity. The existence of this degeneracy demonstrates that
the total velocities of the Be stars are well constrained and that the
model is simply changing the fraction of these velocities arising from
either the velocity dispersion of the disc or the runaway velocity.

 \begin{figure}
 	\begin{subfigure}[t]{\columnwidth}
 		\includegraphics[scale=0.55,trim = 4mm 4mm 0mm 3mm, clip]{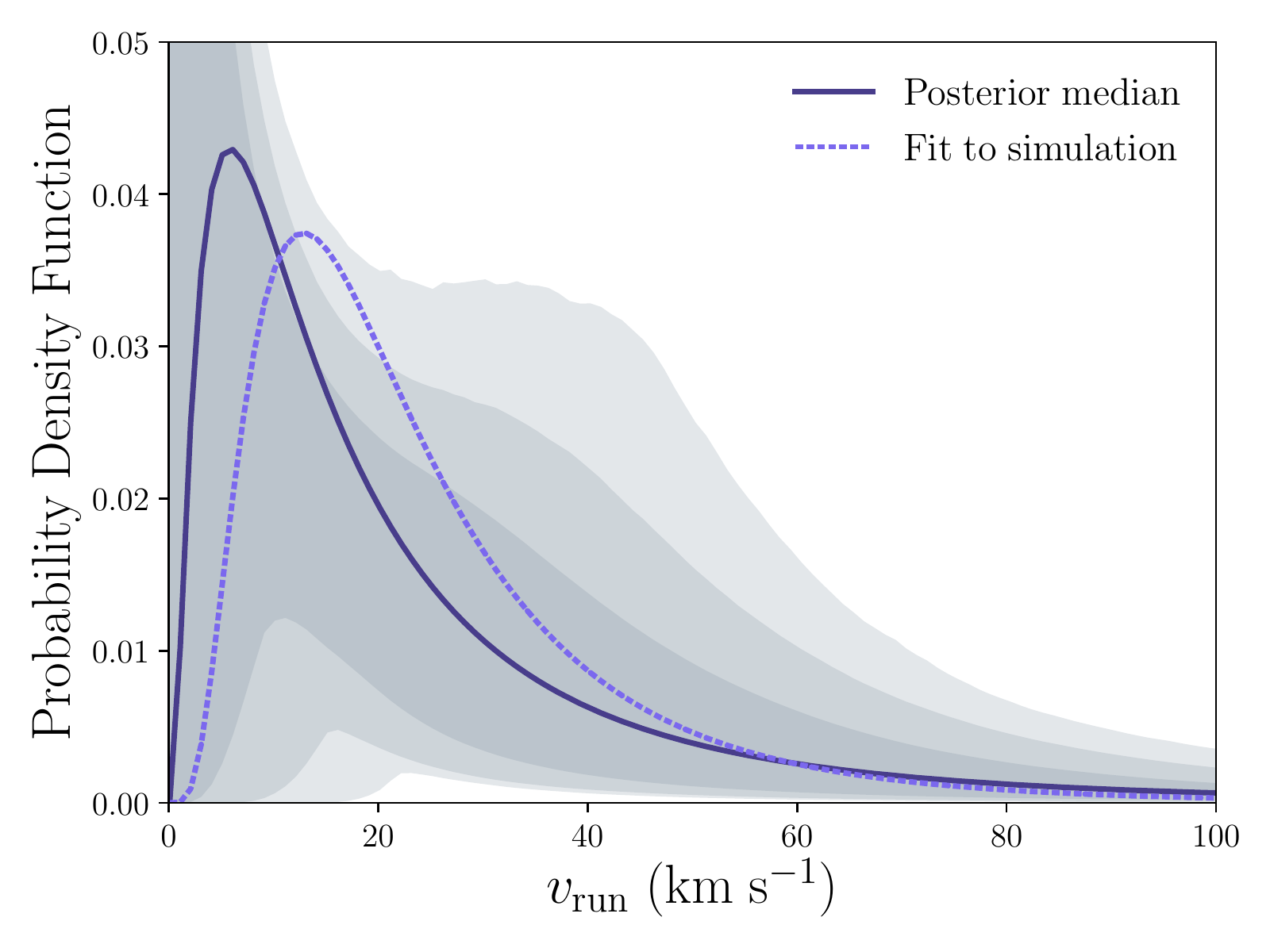}
 		\subcaption{Probability density functions.}
 	\end{subfigure}
 \\ 
	\begin{subfigure}[t]{\columnwidth}
		\vspace{0.3cm}
		\includegraphics[scale=0.55,trim = 4mm 4mm 0mm 3mm, clip]{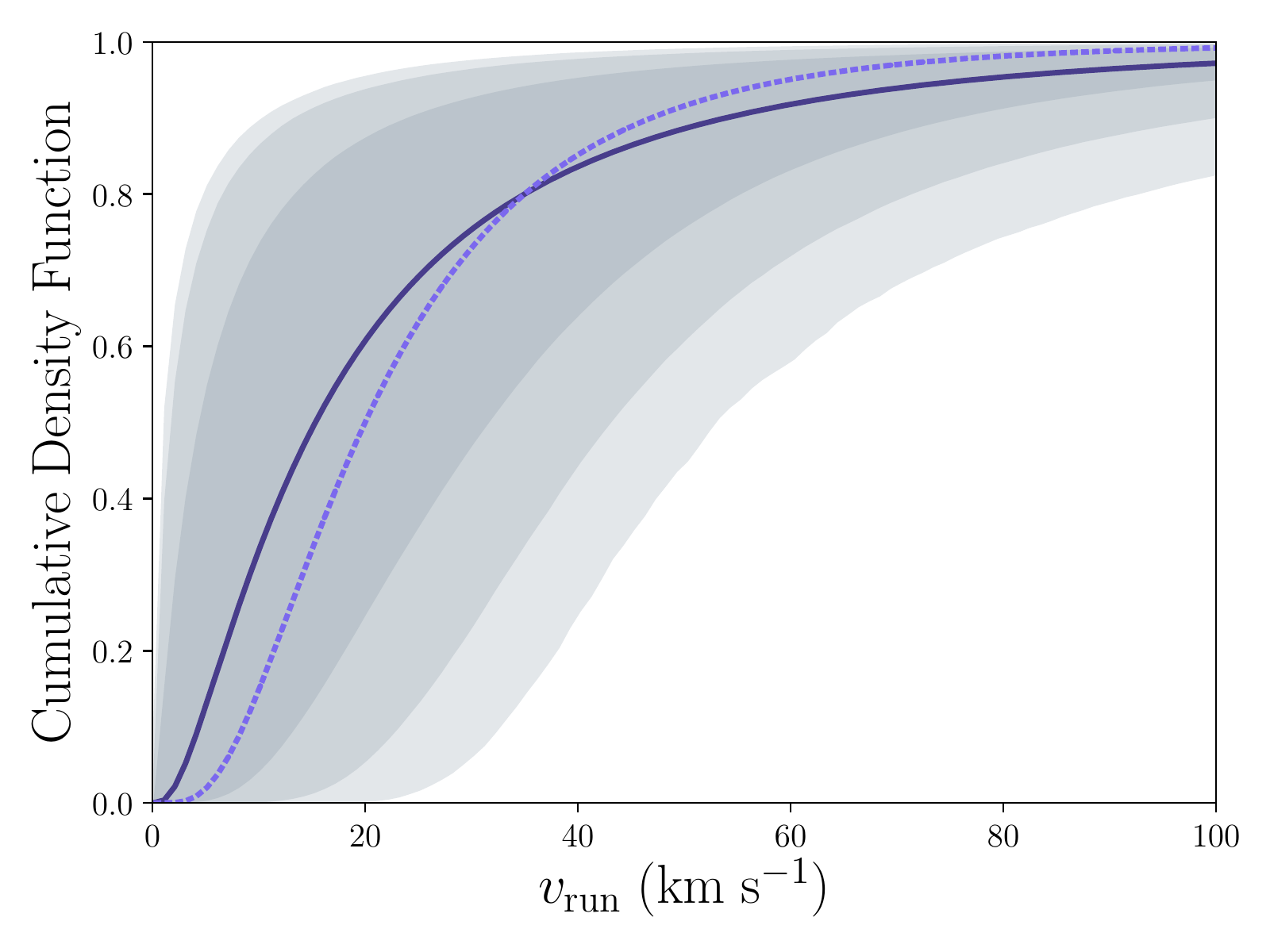}
		\subcaption{Cumulative density functions.}
	\end{subfigure}
	\caption{In grey, we show the 1, 2, and 3$\sigma$ contours of the log-normal distributions implied by the posterior of $(\mu,\sigma)$ shown in
          Fig.~\ref{fig:invertedfree}. The dark line is the log-normal given by the median of the posterior and the purple dashed line is
          the single log-normal fit to the simulated
          runaway velocity distribution shown in green in
          Fig.~\ref{fig:vejfit}.}
	\label{fig:invertedfreedist}
\end{figure}

Each sampled point in the posterior corresponds to a different runaway
velocity distribution. We take these velocity distributions and
compute the 1, 2, and 3$\sigma$ contours of the probability and
cumulative density functions at each value of $v_{\mathrm{run}}$, and
compare in Fig.~\ref{fig:invertedfreedist} to the single log-normal
runaway velocity distribution shown in Fig.~\ref{fig:vejfit}. The fit
single log-normal is not an outlier in these plots which affirms that
the runaway velocity distribution predicted by our simulations of
binary stars is credible.

\section{Discussion}
We recall that, as shown in Sec.~\ref{sec:binaryc}, a fraction of
$17.5\%$ being runaways is consistent with all classical Be stars
originating through mass transfer in binaries. This result arises
because there is a substantial fraction of Be stars which remain bound
post-supernova or avoid the supernova phase entirely. At
$13.1\substack{+2.6 \\ -2.4}\%$, the posterior runaway fraction
obtained in Section \ref{sec:principal} is two standard deviations
below this. We discuss below several possible explanations for this
discrepancy.

\subsection{Large uncertainty and wide velocity dispersion priors}
Our weak prior on the global velocity dispersions may allow higher
  values to be taken at the expense of the number of runaways. Using
  the relations between age and velocity dispersions of
  \citet{aumer_kinematics_2009}, we find that the posterior velocity
  dispersions imply that the population is around $200\;\mathrm{Myr}$
  old, while the Be stars in our model population (accounting for selection effects) have a median age of $35.4\;\mathrm{Myr}$. It
appears likely that the velocity dispersions are overestimated, with
the caveat that it is likely that the \citet{aumer_kinematics_2009}
velocity dispersions for young stars may be driven by their assumed
functional form of the velocity dispersions with age,
\begin{equation}
\sigma(\tau)=v_{10}\left(\frac{\tau+\tau_1}{10\;\mathrm{Gyr}+\tau_1}\right)^{\beta}
\end{equation}
where $v_{10}$ and $\tau_1$ characterize the velocity dispersion at
$10\;\mathrm{Gyr}$ and at birth and $\beta$ describes the efficiency
of stochastic acceleration. The degeneracy between the velocity
  dispersions and the number of runaways could be broken either by
  measuring the velocity dispersion of a population of runaway-free
  stars coeval to the Be stars or by exploiting the precise astrometry
  in Gaia DR2.

\subsection{Uncertainty in binary stellar evolution}

An alternative origin for the uncertainty is that the binary
  evolution simulations rely on a number of uncertain ingredients,
  some of which we discussed in Section \ref{sec:uncertainty} such as
  the natal kick distribution of compact objects, the fallback of
  material onto black holes and common envelope evolution. The use of
  different prescriptions for these aspects could easily resolve the
  discrepancy either by decreasing the frequency of runaway Be stars
  or shifting the runaway velocity distribution to smaller velocities
  (thus increasing the confusion between true runaways and fast disc
  stars).

More subtle changes to the binary evolution simulations could also
  account for part of the difference. For instance, if the
  prescription for tidal locking requires that the two stars need to
  be closer than is the case in reality, then the runaway velocity
  distribution of Be stars will be inflated to faster velocities. The
  details of the simulated population of Be stars are also sensitive
  to prescriptions describing mass transfer and rotation, both of
  which contain their own uncertainties.

Studies of single and binary stars using Gaia data over the next
  few years will drastically constrain these uncertainties in stellar
  evolution, which in turn will allow for a much keener inference to
  be made on the runaway fraction among Be stars.

\subsection{Observational bias against runaway stars}
\label{sec:zobsbias}

Be stars are young and thus are typically found at low Galactic
  latitudes, because they do not live long enough for their vertical
  velocity to be excited through disc heating. However, runaway Be
  stars ejected out of the disc could have a large vertical velocity
  component and thus be found at high latitudes. There will be a trend
  in which high runaway velocity objects are more common at high
  latitudes. Considering the 40 high-likelihood runaway stars
  identified in Table \ref{tab:runaway}, which are necessarily the
  fastest runaway stars in the catalogue, we see that all but two are
  within $400\;\mathrm{pc}$ of the Galactic plane. It is thus possible
  that our observational bias towards nearby objects may exclude some
  of the higher velocity Be runaway stars. If the discrepancy were to
  be entirely explained by this bias, we would need to miss the
  furthest 25\% of Be stars from the disc.

To investigate this possibility, we sample 10,000 Be runaway stars with
velocity dispersions given by the posterior in Section
\ref{sec:principal} and runaway velocities sampled from the
model distribution obtained in Section
\ref{sec:obsbias}. We assume the Be stars are born proportionally to
the disc density found by \citet{bovy_direct_2013}, noting that with a
scale-height of $400\;\mathrm{pc}$ this is likely to over-estimate the
number of high-latitude Be stars. The orbits of the sampled Be stars
are then followed through the Milky Way using \textsc{Galpy} and the
included MWPotential2014 potential \citep{bovy_galpy:_2015}. Even with
this vertically extended distribution of Be star birth locations, we
find that only 20\% of runaway Be stars should be found above
$|z|=0.5\;\mathrm{kpc}$. Using a more realistic vertical dispersion of
$100\;\mathrm{pc}$ derived from our observed sample of non-runaway
stars, we find that only 3\% of Be runaways should be found above
$|z|=0.5\;\mathrm{kpc}$ and only 0.6\% above $|z|=1.0\;\mathrm{kpc}$.

While this bias is not able to resolve the discrepancy, it does answer
a long-standing open question which we discuss in the following
subsection; \citet{martin_origins_2006} found that there were no Be
stars among their 31 high-latitude ($|z|\gtrsim 1.0\;\mathrm{kpc}$) B
runaway stars.

\subsection{Should we find Be stars at high latitude?}

\citet{martin_origins_2006} investigated the origin and evolutionary
status of 48 B stars found far from the plane of the Milky Way and,
surprisingly, found no Be stars in the sample despite the expectation
of at least 10 Be stars based on the incidence of Be stars observed in
the field by \citet{zorec_critical_1997}. Previously,
\citet{slettebak_spectroscopic_1997} identified 8 Be stars between
$0.2 < \lvert z \rvert < 0.9\;\mathrm{kpc}$ from the plane, which
implies only a small overlap with the \citet{martin_origins_2006}
sample covering $0.5<\lvert z \rvert <
2.0\;\mathrm{kpc}$. \citet{martin_origins_2006} provides several
arguments that could explain this absence:

\begin{enumerate}
	\item \label{item:baseline} The short baseline of the data
          may have missed temporarily inactive Be stars.
	\item \label{item:field} Magnitude-limited field studies are
          biased towards younger, brighter stars which have a higher
          rate of Be stars, i.e. the Be stars aren't missing in this
          sample, merely overcounted in the field.
	\item \label{item:cluster} Based on the $v\sin i$ distribution
          and the lack of observed binaries among the dataset, the B stars may be
          mostly dynamically ejected from a cluster environment. It
          has been observed that there are fewer Be stars in clusters
          (although this may be a selection effect).
\end{enumerate}

This list of solutions doesn't consider the possibility that the Be
phenomenon might be correlated with the runaway velocity, which could
naturally explain the observations if very few Be stars are ejected at
high enough velocities to reach high altitudes above the disc. This
kinematic selection effect is mentioned by \citet{martin_origins_2006}
who found that no star ejected with a velocity less than
$50\;\mathrm{km}\;\mathrm{s}^{-1}$ can make it to $0.5\;\mathrm{kpc}$
above the disc even if the ejection velocity is aligned with the
vertical. As we discussed in Section \ref{sec:zobsbias}, only 3\% of
Be runaway stars should be found more than $0.5\;\mathrm{kpc}$ from
the disc according to our predicted, selection-effect-included runaway
velocity distribution, potentially contributing substantially to the
solution of this problem.

\section{Conclusions}
\label{sec:conclusions}

In this work, we have shown that the first Gaia data release,
specifically the Tycho-Gaia Astrometric Solution (TGAS), advances our
understanding of the kinematics of Be stars. With future data releases
of the full Gaia catalogue, we can expect to have a census of
thousands of Be stars with precise kinematics.

We first constructed the largest catalogue of Be stars with full
six-dimensional kinematics to date. This used three sources -- namely,
the \citet{berger_search_2001} survey, the Be Star Spectra database
(BeSS) and the catalogue of \citet{hou_catalog_2016} from LAMOST --
crossmatching with TGAS where necessary to obtain updated proper
motions and parallaxes. Our final combined catalogue contains 632 Be
stars.

We then modelled the evolution of binaries across a grid in parameter
space using the fast {\sc binary\_c} code \citep[see
  e.g.,][]{izzard_population_2009} with a view to testing the
post-mass-transfer model of Be stars~\citep{pols_formation_1991}.  We
computed the probability distribution for the runaway velocity and
critical equatorial velocity ratio.  We developed a criterion based on
the equatorial velocity ratio for Be stars and thus obtained the
distribution of runaway velocities and a prediction that 5\% of all Be stars should be runaways (lower curve in Fig. \ref{fig:runfracduetobias}).
We then demonstrated that it
is vital to account for the observational bias towards early-type Be
stars in our catalogue, because this changes our prediction for the
population. For instance, we find that a fraction of $17.5\%$ Be stars
in our catalogue being runaways is consistent with all classical Be
stars originating through mass transfer in binaries, as the rest
remain bound post-supernova or miss the supernova phase entirely.

To describe the kinematics of Be stars in our sample, we developed a
Bayesian mixture model comprising three populations -- thin disk
stars, runaways and contaminants.  The Bayesian model contained a
total of 5067 parameters to represent the kinematics of 632 Be
stars. Successfully tackling a Bayesian problem of this scope is a
central achievement of this work. It allows us to go beyond fitting a
runaway velocity distribution by modelling the distance, velocity
dispersions and runaway fraction and velocities simultaneously. We
verified that our method is both able to accurately retrieve the input
parameters of an artificially-generated test set of stars and
individually identify with low contamination the fastest third of the
runaway stars. Applying to the true dataset, the posterior runaway
fraction of Be stars is $13.1^{+2.6}_{-2.4}\%$. This suggests that
some Be stars may originate through a process other than the
post-mass-transfer scenario. However, caution is needed as there are a
number of factors that may have caused us to underestimate the runaway
fraction. In particular, there is a degeneracy between velocity
dispersions of the thin disk stars and runaway fraction, and so a weak
prior on the velocity dispersions may allow them to be overestimated
at the cost of a low runaway fraction. The degeneracy between the
velocity dispersions and number of runaway stars will be resolved by
the precise astrometry in Gaia DR2. Additionally, there are a number
of ingredients in binary stellar evolution (such as common envelope
evolution and the distribution of compact object natal kicks) which
are poorly understood and may be contributing to the discrepancy. We
thus conclude that the predicted $17.5\%$ incidence of runaway stars
among Be stars is consistent with our measured fraction and thus all
Be stars could originate through the post-mass-transfer channel.

We also studied the expected distribution of Be stars at high Galactic
latitude.  We have argued that the dearth of Be stars in the high
latitude runaway B star sample of \citet{martin_origins_2006} can be
explained even if most of the runaways originate with the binary
supernova scenario, because we would not expect runaway Be stars to
have high enough velocities to reach high Galactic latitudes.

Although we have not fully resolved the contribution of the
post-mass-transfer channel to the Be star population with the TGAS
data, we are optimistic about the road ahead. The second and later
  Gaia data releases will increase the sample of Be stars with
  accurate kinematics by at least an order of magnitude; for instance,
  \citet{hou_catalog_2016} present a catalogue of 5187 Be stars in
  LAMOST, all of which will have proper-motions and parallaxes in Gaia
  DR2. The Bayesian approach developed in this work to tackle the
kinematics of a large population lays the foundation for the full
exploitation of this future dataset.

\section*{Acknowledgements}
DB is grateful the Science and Technology Facilities Council for
funding his PhD. The authors thank Robert Izzard, Jason Sanders and
the anonymous referee for comments which contributed to the
development of this work. The authors thank the many contributors to
the development of {\sc binary\_c}. The construction of our Be star
catalogue relied on the BeSS database, operated at LESIA, Observatoire
de Meudon, France \citep{neiner_be_2011} and the SIMBAD database,
operated at CDS, Strasbourg, France \citep{wenger_simbad_2000}. This
work made use of the {\sc Python} modules {\sc AstroPy}
\citep{astropy_collaboration_astropy:_2013}, {\sc corner.py}
\citep{foreman-mackey_corner.py:_2016}, {\sc Matplotlib}
\citep{hunter_matplotlib:_2007}, {\sc NumPy} \citep{walt_numpy_2011},
      {\sc Pandas} \citep{mckinney_data_2010} and {\sc SciPy}
      \citep{jones_scipy:_2001}. The inference in this work was made
      possible by the Bayesian inference platform {\sc Stan}
      \citep{carpenter_stan_2017}.

%
%
%
\bibliographystyle{mnras}
\bibliography{references} 
%
%





\bsp	
\label{lastpage}
\end{document}